\documentclass[11pt,a4paper]{article}
\usepackage{jheppub}
\usepackage{amsthm,bm,mathrsfs,stmaryrd}

\newcommand{\be}{\begin{equation}}
\newcommand{\ee}{\end{equation}}
\newcommand{\ba}{\begin{eqnarray}}
\newcommand{\ea}{\end{eqnarray}}

\newcommand{\mt}[1]{$\mathop{#1}$}

\newcommand{\st}{\scriptstyle}
\newcommand{\sst}{\scriptscriptstyle}

\newcommand{\nn}{\nonumber\\}
\newcommand{\eq}{&=&}

\newcommand{\edf}{&:=&}

\def\d{\delta}

\def\e{\epsilon}

\def\cA{{\cal A}}
\def\cB{{\cal B}}
\def\cC{{\cal C}}
\def\cD{{\cal D}}
\def\cE{{\cal E}}

\def\cG{{\cal G}}
\def\cH{{\cal H}}

\def\cK{{\cal K}}
\def\cL{{\cal L}}

\def\cT{{\cal T}}

\def\cV{{\cal V}}

\def\cY{{\cal Y}}
\def\cZ{{\cal Z}}

\author[a]{Euihun JOUNG}

\author[a]{Massimo TARONNA}

\affiliation[a]{Scuola Normale Superiore and INFN\\
Piazza dei Cavalieri 7, 56126 Pisa, Italy}

\emailAdd{euihun.joung@sns.it}
\emailAdd{massimo.taronna@sns.it}

\title{\center Cubic interactions of
massless higher spins\\
 in (A)dS: metric-like approach}

\abstract{
Cubic interactions of higher-spin gauge fields in $(A)dS_{d}$ are studied in the metric-like approach. Making use of the traceless and transverse constraints together with the ambient-space formalism, all consistent parity-invariant cubic vertices
are obtained  for $d\ge4$
in a closed form pointing out the key role of their flat-space counterparts.
}

\begin{document}

\maketitle

\section{Introduction}
\label{sec:intro}

Understanding the systematics of higher-spin (HS) gauge theories\footnote{
For some recent reviews of HS gauge theories,
see e.g. the proceeding \cite{SolvayHS} (which includes contributions \cite{
Bianchi:2005yh, Francia:2005bv, Bouatta:2004kk, Bekaert:2005vh, Sagnotti:2005ns})
and \cite{Sorokin:2004ie, Francia:2006hp,Bekaert:2010hw}.
} has been attracting an increasing attention in recent years, and finding a consistent Lagrangian that describes their interactions is
one of the main problems in the subject.
Vasiliev's equations \cite{Vasiliev:1988sa,Vasiliev:2003ev} provide at present the only known fully non-linear consistent description, at least at the classical level, of an infinite number of HS gauge fields of all spins.\footnote{
See e.g. \cite{Vasiliev:2004qz, Bekaert:2005vh} for some reviews of Vasiliev's equations,
and \cite{Boulanger:2011dd, Colombo:2010fu, Sezgin:2011hq,
Didenko:2008va,Iazeolla:2008bp,Iazeolla:2011cb} for a recent proposal on the action principle, observables and some exact solutions of Vasiliev's equations.} However the nature of their couplings still leaves interesting questions to be answered.
The generalization of the lower-spin gauge interactions of Yang-Mills and Gravity to HS
is associated with a non-linear deformation of the Abelian HS gauge symmetries
of the free theory \cite{Fronsdal:1978rb,Fang:1978wz}\footnote{See \cite{Francia:2002aa,Francia:2002pt, Francia:2005bu, Campoleoni:2008jq, Campoleoni:2009gs,
Campoleoni:2009hj, Campoleoni:2009je}
for the unconstrained formulation of HS gauge theory,
and \cite{Francia:2010ap, Francia:2010qp,Francia:2011qa} for recent developments.
}
and can be studied perturbatively by means of the Noether procedure. This actually rests on enforcing gauge invariance of the full Lagrangian order by order in the number of fields, and has been considered by mainly two different perspectives: frame-like or metric-like formalisms.

Important progress on HS cubic interactions
in the frame-like approach was obtained by Fradkin and Vasiliev (FV) \cite{Fradkin:1986qy,Fradkin:1987ks} who extended the gravitational minimal coupling to \mt{s_{1}\!-\!s_{2}\!-\!s_{3}} HS couplings.
Their construction of cubic couplings
is consistent in (Anti) de Sitter ((A)dS) backgrounds,
and one of its essential features is the presence of inverse powers of the cosmological constant.
Very recently, these interaction vertices in $AdS_{4}$ were generalized to $AdS_{d}$ \cite{Vasiliev:2011xf}\,,
that were conjectured to cover all vertices that can be constructed in terms of connection one-forms and curvature two-forms of symmetric HS gauge fields.
In fact, the goal of the present paper is the same as that of \cite{Vasiliev:2011xf}, and it would be
in principle interesting to explore the relation of our results with those of \cite{Vasiliev:2011xf}.
This comparison is although non-trivial since the two constructions use very different mathematical devices, and we
will only discuss in the conclusion how the FV structure of the vertices is recovered in our approach.
See \cite{Alkalaev:2010af, Zinoviev:2010cr, Boulanger:2011qt, Boulanger:2011se}
for other recent developments in the frame-like approach to the cubic-interaction problem.

On the other hand, the flat-space cubic vertices of HS gauge fields in the metric-like formalism were investigated first by Berends, Burgers and van Dam
\cite{Berends:1984rq,Berends:1985xx}, and then by many other authors.\footnote{
See e.g. the review \cite{Bekaert:2010hw} for an exhaustive list of works.}
Notably, the consistent vertices were classified by Metsaev \cite{Fradkin:1995xy, Metsaev:2005ar, Metsaev:2007rn}
in terms of the number of derivatives within the lightcone approach.
Despite various efforts made along the years by a number of authors, only recently has it been possible to arrive at a covariant description of all bosonic flat-space cubic interactions
by Manvelyan, Mkrtchyan and Ruehl in \cite{Manvelyan:2010wp, Manvelyan:2010jr,Mkrtchyan:2010pp} from a field theoretical perspective.
At the same time,
starting from a String Theory vantage point and with a careful analysis of the gauge invariant pieces contained in the string amplitudes,
all consistent cubic interactions involving any bosonic and fermionic fields
were obtained by Sagnotti and one of the authors in \cite{Taronna:2010qq,Sagnotti:2010at}. These results pointed out the key role of \emph{on-shell} expressions
(the part of vertex that does not involve divergences and traces\footnote{The traceless constraint is needed for the irreducibility of the representations.
For a more detailed analysis of
other possibilities,
see \cite{Taronna:2011kt}.}
of fields)
of the cubic interactions from which one can recursively reconstruct any consistent
\emph{off-shell} cubic action.
Further results and developments on higher-order vertices and scattering amplitudes
can be found in \cite{Fotopoulos:2010ay} and \cite{Taronna:2011kt}.

The metric-like approach to (A)dS cubic vertices was also explored by some authors. For instance,
the $3\!-\!3\!-\!2$ vertex was obtained in
\cite{Zinoviev:2008ck,Boulanger:2008tg},
and the $s\!-\!0\!-\!0$ vertices were constructed in \cite{Manvelyan:2009tf} by using the standard Noether procedure.
The latter were also studied in \cite{Fotopoulos:2007yq}
by using the BRST technique\footnote{
See also \cite{Buchbinder:2006eq} for the general strategy of the construction,
and see \cite{Fotopoulos:2008ka} for a review of the BRST approach.},
and also using current couplings in \cite{Fotopoulos:2010nj, Bekaert:2010hk}.\footnote{
See \cite{Francia:2007qt, Sagnotti:2010jt} for HS current exchanges,
and see \cite{Bekaert:2009ud, Bekaert:2010ky} for flat-space current-coupling interactions
and their application to the effective action in a HS background.
}

\subsection*{Summary of results}

The aim of the present paper is to construct and classify
the consistent parity-invariant cubic interactions of bosonic symmetric HS gauge fields in (A)dS backgrounds of any dimension greater than three
 within the metric-like formalism \cite{Fronsdal:1978vb}.
The expressions of flat-space cubic vertices in this formalism are highly involved\footnote{
See however \cite{Sagnotti:2010at, Taronna:2011kt} and \cite{Manvelyan:2010je,Mkrtchyan:2010pp} for more compact expressions.},
and one can expect that the (A)dS vertices may have even more complicated expressions.
In order to circumvent the complexity, we first exploit the simplicity of working with
the transverse and traceless (TT) constraints on the fields, which we expect to be
systematically removable.
Second, we employ the ambient-space description of (A)dS fields.

\paragraph{Transverse and Traceless constraints}

One of the main lessons in the recent construction of flat-space cubic interactions is that
the full (off-shell) expressions of the vertices are fixed by their on-shell forms.
The latter may be regarded as the consistent cubic-interaction vertices for the system of
HS gauge fields with the TT constraints\footnote{ For completeness, let us mention that the TT \emph{decomposition} has been introduced, although from a slightly different perspective, in the
$s=2$ case \cite{Deser:1959zza}.}. In this approach the kinetic term of HS fields $\varphi_{\mu_{1}\cdots \mu_{s}}$ becomes simply $\varphi^{\,\mu_{1}\cdots \mu_{s}}\,\partial^{2}\,\varphi_{\mu_{1}\cdots \mu_{s}}$\,, and is invariant under \mt{\delta\, \varphi_{\mu_{1}\cdots \mu_{s}}= \partial_{(\mu_{1}} \varepsilon_{\mu_{2}\cdots \mu_{s})}} with the gauge parameters $\varepsilon_{\mu_{1}\cdots \mu_{s-1}}$ also subjected to TT constraints as well as an additional differential constraint \mt{\partial^{2}\,\varepsilon_{\mu_{1}\cdots \mu_{s-1}}=0}\,.
The key observation is
that the consistent cubic-interaction problem can be addressed at this level,
while foregoing the constraints requires a tedious but well defined procedure.
Therefore, also for the (A)dS cubic interactions, we work within the TT setting where the transverse constraint is now with respect to the (A)dS-covariant derivative.

\paragraph{Ambient-space formalism}
Differently from the flat-space vertices, even after imposing the TT constraints
the structure of (A)dS cubic vertices is still highly non-trivial due to the non-commutativity of the covariant derivatives. A further simplification is achieved
making use of the ambient-space formalism \cite{Fronsdal:1978vb,
Metsaev:1995re,Metsaev:1997nj, Biswas:2002nk} of (A)dS fields,
where one rewrites intrinsic (A)dS quantities in terms of simpler flat-space ones.\footnote{
The ambient-space formalism has been used for a large number of applications. See e.g.
\cite{Bekaert:2003uc, Hallowell:2005np,Barnich:2006pc,Fotopoulos:2006ci,Francia:2008hd,Boulanger:2008up,Boulanger:2008kw,Alkalaev:2009vm}.}
Recently, it was also exploited in order to construct spin-$s$ gauge interactions with a scalar field \cite{Bekaert:2010hk}.

The key feature of the ambient-space formalism is to regard the (A)dS space as the codimension-one hyper-surface \mt{X^{2}=L^{2}} in an ambient flat space-time parameterized by Cartesian coordinates $X^{\sst M}$ with ${\st M}=0, 1, \cdots, d$\,. In this formalism, the ambient-space HS fields $\Phi_{\sst M_{1}\cdots M_{s}}$ that are homogeneous in $X^{\sst M}$ and tangent to the hyper-surface, are in one-to-one correspondence to the (A)dS fields $\varphi_{\mu_{1}\cdots \mu_{s}}$\,. Moreover, the field equations and gauge transformations, as well as the TT constraints of (A)dS fields, can be derived from those of the ambient-space fields by a radial-dimensional reduction. The only subtlety of this formalism arises from the formally diverging radial integral at the level of the action.
This can be cured with a $\d$-function insertion of the form
 $\delta\big(\tfrac{\sqrt{X^{2}}}L-1\big)$\,. The presence of $\delta$-function is the main difference between the flat-space constructions and the ambient-space (A)dS ones.  It requires particular care since it spoils the usual flat-space property that the integral of a total derivative vanishes.

\medskip

With the aid of the TT constraints and the ambient-space formalism, the problem of finding consistent cubic interactions of HS fields in (A)dS becomes almost the same
problem of flat-space vertices.
The only difference, as we anticipated, is that the ambient-space action contains a $\d$-function
insertion, which makes  the ambient-space total-derivative terms arising from gauge variations non-trivial.
In order to compensate these gauge-variation terms, cubic-interaction vertices
must also include proper total-derivative terms.
To summarize,
all consistent (A)dS cubic interactions read
\ba
\label{sol ads}
	&& S^{\sst (3)} = \frac{1}{3!} \sum_{s_{1},s_{2},s_{3}=0}^{\infty}
	\sum_{n=0}^{\rm min\{s_{1},s_{2},s_{3}\}} g^{s_{1}s_{2}s_{3},n}_{a_{1}a_{2}a_{3}}\,
	\int \frac{d^{d+1}X}L\,\delta\Big(\tfrac{\sqrt{X^{2}}}L-1\Big)\,\times \nn
	&&\times\,
	\big[\partial_{U_{1}}\!\!\cdot(\partial_{X_{23}}\!\!+\alpha\,\partial_{X})\big]^{s_{1}-n}\,
	\big[\!-2\,\partial_{U_{2}}\!\!\cdot(\partial_{X_{1}}\!\!
	-\tfrac{\alpha-1}{\alpha+1}\,\partial_{X})\big]^{s_{2}-n}\,
	\big[\,2\,\partial_{U_{3}}\!\!\cdot(\partial_{X_{1}}\!\!-\tfrac{\alpha+1}{\alpha-1}\,\partial_{X})\big]^{s_{3}-n}
	\nn
	&& \times \,
	\big[ \partial_{U_{2}}\!\!\cdot\partial_{U_{3}}\,
	\partial_{U_{1}}\!\!\cdot(\partial_{X_{23}}\!\!+\beta\,\partial_{X})
	-2\,\partial_{U_{3}}\!\!\cdot\partial_{U_{1}}\,
	\partial_{U_{2}}\!\!\cdot(\partial_{X_{1}}\!\!+\tfrac{\alpha-\beta}{\alpha+1}\,\partial_{X}) \\	
	&& \quad\ \,+\, 2\,\partial_{U_{1}}\!\!\cdot\partial_{U_{2}}\,
	\partial_{U_{3}}\!\!\cdot(\partial_{X_{1}}\!\!+\tfrac{\alpha-\beta}{\alpha-1}\,\partial_{X})
	\big]^{n}\
	\Phi^{a_{1}}(X_{\sst 1},U_{\sst 1})\
	\Phi^{a_{2}}(X_{\sst 2},U_{\sst 2})\ \Phi^{a_{3}}(X_{\sst 3},U_{\sst 3})\,
	\Big|_{\overset{X_{1}=X_{2}=X_{3}=X}{\sst U_{1}=U_{2}=U_{3}=0}}\,,
	\nonumber
\ea
where the form of the vertices is encoded in a differential operator acting
on the generating function of ambient-space fields:
\be
\Phi^{\sst a}(X,U)\,=\,\sum_{s\,=\,0}^\infty\tfrac{1}{s!}\
\Phi^{\sst a}_{\sst M_{1}\cdots M_s}(X)\,U^{\sst M_1}\cdots U^{\sst M_s}\,,
\ee
while
 \mt{\partial_{X^{\sst M}}=\partial_{X_{1}^{\sst M}}+\partial_{X_{2}^{\sst M}}+\partial_{X_{3}^{\sst M}}} is the total derivative,
 and \mt{\partial_{X_{ij}^{\sst M}}:=\partial_{X_{i}^{\sst M}}-\partial_{X_{j}^{\sst M}}}\,.
Different choices of $\alpha$ and $\beta$ can be in fact absorbed into the coupling constants, and hence one can work with any particular choice.
The number of \emph{ambient-space derivatives} in \eqref{sol ads} is
\be
\Delta=s_{1}+s_{2}+s_{3}-2n\,,
\ee
but, when radially reduced, different pieces of the (A)dS vertices
involve different number of covariant derivatives: \mt{\Delta,\ \Delta-2,\, \dots\,, 1} (or $0$), while whenever the number of derivatives decreases by two the corresponding mass-dimension is compensated by the cosmological constant $\Lambda:=1/L^{2}$\,.\footnote{The correct relation between the cosmological constant $\Lambda_{\sst\rm C.C.}$ and the radius of (A)dS is \mt{\Lambda_{\sst\rm C.C.}=(d-1)(d-2)/(2\,L^{2})=\Lambda\,(d-1)(d-2)/2}\,. However in this paper, for simplicity, we call also $\Lambda$ cosmological constant.}
This structure makes clear the relation of \eqref{sol ads} to the FV vertices, where the inverse-power expansion in $\Lambda$\, appears. For instance, concentrating on the gravitational couplings (\mt{s_{1}=s_{2}=s} and \mt{s_{3}=n=2}) in \eqref{sol ads},
the action can be recast in terms of an inverse-power series in $\Lambda$ as
\be
	S^{\sst (2)}+S^{\sst (3)} =  \frac{\lambda_{s}}G\ 	\sum_{r=2}^{s}\,
	\frac{1}{\Lambda^{r-2}}\,\int_{(A)dS_{d}}\cL_{r}\,.
	 \label{act ss2}
\ee
In order to get this expression,
we made use of the redefinitions \mt{g^{ss{\sst 2,2}}=\Lambda^{2-s}\sqrt{G}\ \lambda_{s}} and \mt{\varphi^{\sst (s)}=\phi^{\sst (s)}/\sqrt{G}}
with the gravitational coupling constant $G$\,.
$\cL_{r}$'s are cubic vertices
which are separately gauge invariant under the spin 2 gauge transformation
and can be written schematically as
\be
	\cL_{r}=D^{2(r-1)}\,h\ \phi^{\sst (s)} \phi^{\sst (s)}
	+\Lambda\,D^{2(r-2)}\,h\ \phi^{\sst (s)} \phi^{\sst (s)}\,,
	\label{FV ver}
\ee
where  $\cL_{2}$ should involve the gravitational minimal coupling.
Notice as well that
the inverse-power $\Lambda$-expansion has its origin from the redefinition of the coupling constant $g^{ss{\sst 2,2}}$\,, which makes the two-derivative part of the vertex independent of $\Lambda$\,.
A particular flat-space limit of the FV vertices considered by Boulanger, Leclercq and Sundell \cite{Boulanger:2008tg} is also discussed along the same lines.

The outline of this paper is as follows. In Section~\ref{sec: flat cubic}, we review
the consistent flat-space cubic vertices,
which will play an important role in the (A)dS case. In Section~\ref{sec: ads free}, we define the TT-constrained system of free HS gauge fields in (A)dS, and reformulate it in the ambient-space formalism. Section~\ref{sec: ads cubic} is the main part of the paper, where we construct the consistent cubic interactions in (A)dS within
 the TT and ambient-space formalisms. We show also how to get the (A)dS intrinsic expressions by radially reducing the ambient-space ones. Section~\ref{sec: discussions} contains discussions of our results. In Appendix~\ref{sec: forego} we review how to forego the constraints for the case of flat-space vertices and then discuss the (A)dS cases. Finally, Appendices~\ref{sec:C=0} and \ref{sec:reduc} contain some technical
 details on our constructions.


\section{Cubic interactions of HS gauge fields in flat space}
\label{sec: flat cubic}

Let us begin with reviewing the recent results on flat-space cubic vertices.
The construction of on-shell cubic vertices is relatively simple, and
the corresponding  \emph{off-shell} versions can be systematically recovered.
Notice that
the construction of flat-space cubic vertices with the TT constraints
can be viewed as the covariant version of Metsaev's
lightcone result \cite{Metsaev:2005ar}.

In order to distinctly separate the initial construction and the later completion of vertices,
we slightly change our viewpoint and begin with a system of HS fields subject to the traceless and transverse (TT) constraints. This is nonetheless on its own a consistent setting to search for consistent deformations of the free theory.
In this section, we first formulate the free HS gauge theory as a TT-constrained system
and then construct the consistent cubic interactions for that system.

\subsection{Traceless and Transverse HS gauge fields}

In order to work with compact expressions, we resort to generating functions of symmetric tensor fields,
where the HS fields $\varphi_{\mu_{1}\cdots\mu_{s}}$
are contracted with auxiliary variables $u^{\mu}$ as
\be
	\varphi(x,u):=\sum_{s=0}^{\infty}\frac1{s!}\,\varphi_{\mu_{1}\cdots\mu_{s}}(x)\,
	u^{\mu_{1}}\cdots u^{\mu_{s}}\,.
\ee
In this notation the TT constraints can be written as
\be
	\partial_{u}^{2}\,\varphi=0\,, \qquad \partial_{u}\!\cdot\partial_{x}\,\varphi=0\,.
	\label{tt flat}
\ee
For this constrained system of fields, the quadratic action does not allow any non-trivial tensor contraction and is given by a \emph{scalar-field-like} action of the form:
\be
	S^{\sst (2)} = \frac12\int d^{d}x\,
	\Big[\, \delta_{a_{1}a_{2}}\,e^{\partial_{u_{1}}\!\cdot\,\partial_{u_{2}}}\,
	\varphi^{a_{1}}(x_{1},u_{1})\,\partial_{x_{2}}^{2}\,\varphi^{a_{2}}(x_{2},u_{2})\,
	\Big]_{\overset{x_{i}=x}{\sst u_{i}=0}} \,.
	\label{flat q act}
\ee
Here we have introduced \emph{colors} for HS fields, labeled by $a_{i}$\,, for the description of Yang-Mills-like interactions associated to Chan-Paton factors. It is useful to notice that the field equations for this system simply reads
\be
	\partial_{x}^{2}\,\varphi\approx 0\,,
	\label{eom flat}
\ee
where $\approx$ means herefrom that the equality holds modulo the free field equation. This action is invariant under the usual linearized gauge transformations:
\be
	\delta^{\sst (0)}_{\varepsilon}\,\varphi=u\cdot\partial_{x}\,\varepsilon\,.
	\label{gt flat}
\ee
Due to the TT constraints of the fields, the gauge parameters are constrained as well to satisfy the TT constraints:
\be
	\partial_{u}^{2}\,\varepsilon=0\,, \qquad \partial_{u}\!\cdot\partial_{x}\,\varepsilon=0\,,
	\label{tt gp flat}
\ee
and also one additional field-equation-like differential constraint:
\be
	\partial_{x}^{2}\,\varepsilon=0\,.
	\label{eom gp flat}
\ee
In fact, the latter constraints enable one to forego the TT constraints order by order in the number of divergences and traces.

As one can see, the TT constraints are closely related to the representations of the Lorentz group.
In this respect, the transverse constraint can be considered as the condition to deal with
a general representation of the Lorentz group while the traceless constraint selects irreducible ones.
We keep the TT constraints for gauge fields \eqref{tt flat} and gauge parameters \eqref{tt gp flat} at the interacting level, but the gauge transformations \eqref{gt flat}
and the last constraint of gauge parameters \eqref{eom gp flat} will be deformed by cubic-interaction vertices.

\subsection{General solutions for consistent cubic interactions}

In this section we review the construction of the consistent parity-invariant
cubic interactions for $d\ge4$ that deform the free HS gauge theory.
We begin with the most general expression for a cubic interaction which, in terms of generating functions, can be written in the form:\footnote{
For example,  $\partial^{\nu}\varphi_{\mu_{1}\mu_{2}}\,\partial^{\rho_{1}}
\varphi_{\nu}^{\ \ \mu_{1}\mu_{2}\rho_{2}}\,\varphi_{\rho_{1}\rho_{2}}$
is given by $C=\partial_{x_{1}}\!\!\cdot\partial_{u_{2}}\,\partial_{x_{2}}\!\!\cdot\partial_{u_{3}}
\,(\partial_{u_{1}}\!\!\cdot\partial_{u_{2}})^{2}\,\partial_{u_{2}}\!\!\cdot\partial_{u_{3}}$\,.
}
\ba
	S^{\sst (3)} \eq
	\frac1{3!}\int d^{d}x\
	C_{a_{1}a_{2}a_{3}}(\partial_{x_1},\partial_{x_{2}}, \partial_{x_{3}};
	\partial_{u_1}, \partial_{u_{2}}, \partial_{u_{3}})\times\nn
	&& \hspace{45pt} \times\,
	\varphi^{a_{1}}(x_{1},u_{1})\ \varphi^{a_{2}}(x_{2},u_{2})\
	\varphi^{a_{3}}(x_{3},u_{3})\,\Big|_{\overset{x_{i}=x}{\sst u_{i}=0}}\,,
	\label{f ansz}
\ea
where $C_{a_{1}a_{2}a_{3}}$ satisfies the symmetry properties:
\be
	C_{a_{2}a_{1}a_{3}}(\,2\,,\,1\,,\,3\,)
	=C_{a_{1}a_{3}a_{2}}(\,1\,,\,3\,,\,2\,)
	=C_{a_{1}a_{2}a_{3}}(\,1\,,\,2\,,\,3\,)\,.
	\label{c sym}
\ee
Taking into account integrations by parts and the TT constraints \eqref{tt flat},
 we can conclude that
$C_{a_{1}a_{2}a_{3}}$ depends in principle on
all possible Lorentz invariants
for the parity-invariant interactions\footnote{
One can consider as well the totally antisymmetric tensor $\epsilon_{\mu_{1}\cdots\mu_{d}}$
for the parity-violating interactions, which however do not
exist for $d>7$\,.
See \cite{Boulanger:2005br} for parity-violating cubic vertices of spin 3 fields.} as
\be
	C_{a_{1}a_{2}a_{3}}\big(\,
	\cY_{1}\,,\, \cY_{2}\,,\,\cY_{3}\,,\,
	\cZ_{1}\,,\,\cZ_{2}\,,\,\cZ_{3}\,\big)\,,
	\label{lorentz}
\ee
where
\ba
	&& \cY_{1}:=\partial_{u_{1}}\!\!\cdot\partial_{x_{23}}\,,\quad
	\cY_{2}:=\partial_{u_{2}}\!\!\cdot\partial_{x_{31}}\,,\quad
	\cY_{3}:=\partial_{u_{3}}\!\!\cdot\partial_{x_{12}}\,, \nn
	&& \cZ_{1}:=\partial_{u_{2}}\!\!\cdot\partial_{u_{3}}\,,\quad\
	\cZ_{2}:=\partial_{u_{3}}\!\!\cdot\partial_{u_{1}}\,,\quad\
	\cZ_{3}:=\partial_{u_{1}}\!\!\cdot\partial_{u_{2}}\,.
	\label{Y&Z}
\ea
Here we made use of the compact notation \mt{\partial_{x_{ij}^{\mu}}:=\partial_{x_{i}^{\mu}}-\partial_{x_{j}^{\mu}}}\,.
This choice makes it
possible to get rid of the ambiguity related to the total derivative terms.
Notice that we have excluded terms proportional to $\partial_{x_{i}}\!\cdot\partial_{x_{j}}$'s since they are proportional to the field equations up to a total derivative term,
so that they can be removed by proper field redefinitions.

The Noether procedure, the order by order gauge invariance of the action,
gives at the cubic level
\be
	\delta^{\sst (1)}\,S^{\sst (2)}+\delta^{\sst (0)}\,S^{\sst (3)}=0
	\quad \Rightarrow \quad
	\delta^{\sst (0)}\,S^{\sst (3)} \approx 0\,.
	\label{Noether}
\ee
This, in our notation for $d\ge4$\footnote{
In 3 dimension, one should explicitly take into account that the Weyl tensor vanishes.},
is equivalent to
\be
	\big[\, C_{a_{1}a_{2}a_{3}}\,,\,
	u_{\sst i}\!\cdot\partial_{x_{i}}\,\big]\left(\,\cdots\right)\Big|_{u_i\,=\,0}\approx ({\rm total\ derivative})\,.
	\label{GIR}
\ee
Each $ C_{a_{1}a_{2}a_{3}}$ will define a proper nonlinear deformation
(to the first order in the fields) of the linear gauge symmetries.
Making use of the identities
\be
	\big[\, \cY_{i}\,,\,u_{\sst j}\!\cdot\partial_{x_{j}}\,\big]
	\approx ({\rm total\ derivative})\,, \quad
	\big[\,\cZ_{i}\,,\,u_{\sst j}\!\cdot\partial_{x_{j}}\,\big] =
	-\tfrac12\,\epsilon_{ijk}\,\cY_{k}+
	({\rm total\ derivative}),
	\label{YZ id}
\ee
with  the antisymmetric Levi-Civita symbol  $\epsilon_{ijk}$\,,
the gauge consistency condition \eqref{GIR} gives
the following differential equation for $C_{a_{1}a_{2}a_{3}}$ of \eqref{lorentz}
and its cyclic permutations on $(1)(2)(3)$:
\be
	\left(\cY_{1}\,\partial_{\cZ_{2}}-\cY_{2}\,\partial_{\cZ_{1}}\right)
	C_{a_{1}a_{2}a_{3}}(\,\cY_{i}\,,\,\cZ_{i}\,)=0\,.
	\label{dfe c}
\ee
The general solution for these differential equations
is given by
\be
	C_{a_{1}a_{2}a_{3}}(\,\cY_{1}\,,\,\cY_{2}\,,\,\cY_{3}\,,\,\cZ_{1}\,,\,\cZ_{2}\,,\,\cZ_{3}\,)
	= \cK_{a_{1}a_{2}a_{3}}(\,\cY_{1}\,,\,\cY_{2}\,,\,\cY_{3}\,,\,\cG\,)\,,
	\label{sol flat}
\ee
where the operator $\cG$ is defined as
\ba
	\mathcal G &:=& \cY_{1}\,\cZ_{1}+\cY_{2}\,\cZ_{2}+\cY_{3}\,\cZ_{3} \nn
	&\ =&
	\partial_{u_{2}}\!\!\cdot\partial_{u_{3}}\,\partial_{u_{1}}\!\!\cdot\partial_{x_{23}}
	+\partial_{u_{3}}\!\!\cdot\partial_{u_{1}}\, \partial_{u_{2}}\!\!\cdot\partial_{x_{31}}
	+\partial_{u_{1}}\!\!\cdot\partial_{u_{2}}\,\partial_{u_{3}}\!\!\cdot\partial_{x_{12}}\,.
\ea
Finally, the consistent cubic interactions are encoded in
an arbitrary function $\cK_{a_{1}a_{2}a_{3}}$ \eqref{sol flat}, and when expanded for different spins, they are expressed as
\be
	\cK_{a_{1}a_{2}a_{3}}=
	\sum_{s_{1},s_{2},s_{3}} \!\!\!\!
	\sum_{n=0}^{{\rm min}\{s_{1},s_{2},s_{3}\}}
	g^{s_{1}s_{2}s_{3},n}_{a_{1}a_{2}a_{3}}\ \,
	\mathcal G^{n}\
	\cY_{1}^{s_{1}-n}\,
	\cY_{2}^{s_{2}-n}\,
	\cY_{3}^{s_{3}-n}\,,
	\label{V b}
\ee
where the $g^{s_{1}s_{2}s_{3},n}_{a_{1}a_{2}a_{3}}$'s are coupling constants
that might be fixed by the consistency of higher-order interactions.
The number of derivatives of each vertex can be identified from the above expansion as
\be
	s_{1}+s_{2}+s_{3}-2n\,,
\ee
so that the mass-dimensions of the coupling constants are
\be
	\big[g^{s_{1}s_{2}s_{3},n}_{a_{1}a_{2}a_{3}}\big]
	= \tfrac{6-d}2-s_{1}-s_{2}-s_{3}+2n\,.
	\label{cc md}
\ee
Moreover, from the symmetry properties \eqref{c sym}, the coupling constants inherit the symmetries
\be
	g^{s_{2}s_{1}s_{3},n}_{a_{2}a_{1}a_{3}}=
	g^{s_{1}s_{3}s_{2},n}_{a_{1}a_{3}a_{2}}
	=(-1)^{s_{1}+s_{2}+s_{3}}\,g^{s_{1}s_{2}s_{3},n}_{a_{1}a_{2}a_{3}}\,.
	\label{cc cyc}
\ee
As a result, the uncolored case is consistent only when the
total spin \mt{s_{1}+s_{2}+s_{3}} is even.

\section{Free (A)dS HS gauge fields and Ambient-space formalism}
\label{sec: ads free}

The full expressions for flat-space cubic vertices can be highly involved,
and we have seen that the use of the TT constraints considerably simplifies the problem. Therefore, we employ the same strategy for the construction of cubic interactions in (A)dS,
considering the system of (A)dS HS gauge fields subject to the TT constraints, where the transverse constraint is now with respect to the (A)dS covariant derivative. Again, we expect that
these constraints can be systematically foregone  as in the flat-space case.

However, even after imposing the TT constraints, the (A)dS cubic-interaction problem
continues to present non-trivial difficulties due to the non-commutativity of the covariant derivatives. We simplify the problem one step further making use of the ambient-space formalism of (A)dS fields.

In this section, we will concentrate on the free theory preparing the ground for the construction of cubic vertices which will be the subject of the next section.

\subsection{Traceless and Transverse HS gauge fields}

To begin with, let us introduce the generating functions for symmetric tensor fields $\varphi_{\mu_{1}\cdots\mu_{s}}$ in (A)dS as
\be
	\varphi(x,u)=\sum_{s=0}^{\infty}\frac1{s!}\,\varphi_{\mu_{1}\cdots\mu_{s}}(x)\,
	u\cdot e^{\mu_{1}}(x)\cdots u\cdot e^{\mu_{s}}(x)\,,
\ee
where the fields are contracted with flat auxiliary variables $u^{\alpha}$ via
the inverse vielbein $e^{\ \mu}_{\alpha}(x)$\,: $u\cdot e^{\mu}(x)=u^{\alpha}\,e^{\ \mu}_{\alpha}(x)$\,. We consider, as in the flat-space case, the system of HS gauge fields with traceless and transverse constraints\,:
\be
	\partial_{u}^{2}\,\varphi=0\,,\qquad \partial_{u}\cdot e^{\mu}\,D_{\mu}\,\varphi=0\,,
	\label{tt dS}
\ee
where now the transversality condition is to be considered with respect to the covariant derivative $D_{\mu}$\,, given by
\be
	D_{\mu} := \nabla_{\mu}+\tfrac12\,\omega_{\mu}^{\alpha\beta}(x)
	\,u_{[\alpha}\partial_{u^{\beta]}}\,.
	\label{cov}
\ee
Here the usual (A)dS-covariant derivative $\nabla_{\mu}$ acts on the tensor indices and $\omega_{\mu}^{\alpha\beta}$ is the (A)dS spin connection. With this definition of $D_{\mu}$\,, involving $\nabla_{\mu}$\,instead of $\partial_{\mu}$, the (A)dS Laplacian operator is simply given by $D^{2}$\,.
The quadratic action of HS gauge fields in (A)dS can be then simplified by the TT constraints to
\be
	S^{\,\sst (2)}_{\Lambda}= \frac12\int d^{d}x\sqrt{-g}\,\bigg[
	\delta_{a_{1}a_{2}}\,e^{\partial_{u_{1}}\!\cdot\,\partial_{u_{2}}}\,
	\varphi^{a_{1}}(x_{1},u_{1})
	\left(D^{\,2}_{\sst 2}+\Lambda\,\mu_{2} \right)
 	\varphi^{a_{2}}(x_{2},u_{2})
	\bigg]_{\overset{x_{i}=x}{\sst u_{i}=0}}\,,
	\label{AdS act}
\ee
where $g_{\mu\nu}$ is the (A)dS metric, and $\mu$ is the mass-squared-like operator:
\be
	\mu:=(u\cdot\partial_{u}-2)(u\cdot\partial_{u}+d-4)-2\,,
\ee
while $\Lambda$ is proportional to the cosmological constant, and the subscripts $i$ of $D_{i\,\mu}$ and $\mu_{i}$ indicate that the corresponding operators are acting on $(x_{i}, u_{i})$\,. This action is invariant under the linearized HS gauge transformations:
\be
	\delta^{\sst (0)}_{\varepsilon}\,\varphi=u\cdot e^{\mu}\,D_{\mu}\,\varepsilon\,,
	\label{dS gt}
\ee
with constrained gauge parameters satisfying the TT constraints:
\be
	\partial_{u}^{\,2}\,\varepsilon=0\,, \qquad \partial_{u}\cdot e^{\mu}\,D_{\mu}\,\varepsilon=0\,,
	\label{tt gp dS}
\ee
as well as the field-equation-like differential constraint:
\be
	\left[D^{2}+\Lambda\,u\cdot\partial_{u}
	(u\cdot\partial_{u}+d-2)\right]\varepsilon=0\,.
	\label{eom gp dS}
\ee
One could attempt to construct the cubic interactions at this stage as we did in the flat-space case, but as anticipated, the non-commutative nature of the covariant derivatives makes the problem involved.
To circumvent this problem we choose to work
in the ambient space approach expressing intrinsic (A)dS quantities in term of commutative flat-space ones.

\subsection{Ambient space formalism}

In this subsection we review the ambient-space description for (A)dS HS gauge fields.

\paragraph{Ambient space formalism for dS tensors}

It is well known that one can embed the $d$-dimensional dS space in the $(d+1)$-dimensional flat space with metric:
\be
	ds^{2}_{\rm Amb}=\eta_{\sst MN}\,dX^{\sst M}\,dX^{\sst N}\,,
	\qquad
	\eta=(-,+,\,\cdots,+)\,.
\ee
The dS space is then defined as the hyper-surface $X^{2}=1/\Lambda$ with $\Lambda>0$\,. The AdS space can be obtained from the dS space by an analytic continuation of $\Lambda$\,,\footnote{
One can express AdS tensors in terms of ambient-space ones with a different signature,
but the isomorphism between AdS and ambient-space gauge symmetries is not automatically realized,
contrary to the dS case.}
 and from now on we consider only the dS case with radius $L=1/\sqrt{\Lambda}$\,.

We concentrate on the region of the ambient space with $X^{2}>0$\,, and consider the generating function of symmetric tensor fields $\Phi_{\sst M_{1}\cdots M_{s}}$ given by
\be
	\Phi(X,U)=\sum_{s=0}^{\infty}\frac1{s!}\,
	\Phi_{\sst M_{1}\cdots M_{s}}(X)\,U^{\sst M_{1}}\cdots U^{\sst M_{s}}\,.
\ee
These fields are equivalent to symmetric tensor fields in dS if they are homogeneous in $X^{\sst M}$ and tangent to constant $X^{2}$ surfaces. The latter
conditions translate at the level of the generating function into the following differential constraints:
\be
	(X\cdot\partial_{X}-\Delta_{h})\,\Phi=0\,,
	\qquad
	X\cdot \partial_{U}\,\Phi=0\,,	
	\label{ht con}
\ee
where $\Delta_{h}$ is the degree of homogeneity which will be
fixed later in order to incorporate the gauge symmetries.

In order to identify ambient-space fields with dS fields, it is convenient to first parameterize the $X^{2}>0$ region with the radial coordinates $(R,x)$ given by
\be
	X^{\sst M}=R\,\hat X^{\sst M}(x)\,,\qquad \hat X^{2}(x)=1\,,
\ee
where $R=\sqrt{X^{2}}$ is the radius of dS and the $x^{\mu}$'s are angular variables which also play the role of dS intrinsic coordinates.
The metric in the radial coordinate system is then given by
\be
	ds^{2}_{\rm Amb}=
	dR^{2}+\tfrac{R^{2}}{L^{2}}\,g_{\mu\nu}(x)\,d{x}^{\mu}\,d{x}^{\nu}
	=\tilde e^{\sst R}\,\tilde e^{\sst R}+ \eta_{\alpha\beta}\,\tilde e^{\alpha}\,
	\tilde e^{\beta}\,,
\ee
where $g_{\mu\nu}(x)$ is the dS metric\,, and $(\tilde e^{\sst R},\tilde e^{\alpha})$ is the ambient-space vielbein:
\be
	\tilde e^{\sst R}_{\ {\sst R}}(R,x)=1\,,\quad \tilde e^{\sst R}_{\ \mu}(R,x)=0\,,
	\qquad
	\tilde e^{\alpha}_{\ {\sst R}}(R,x)=0\,,\quad \tilde e^{\alpha}_{\ \mu}(R,x)
	= \tfrac {R}{L}\,e^{\alpha}_{\ \mu}(x)\,.
\ee
For the radial reduction of the generating function of HS fields, it is also necessary to perform for the auxiliary $U^{\sst M}$-variables
the change of coordinates
\be
	U^{\sst M}= \tilde e_{\sst R}^{\ {\sst M}}(R,x)\,v +
	\tilde e_{\sst \alpha}^{\ {\sst M}}(R,x)\,u^{\alpha}
	= \hat X^{\sst M}(x)\,v+ L\,\tfrac{\partial \hat X^{M}}{\partial x^{\mu}}(x)\,
	e_{\alpha}^{\ \mu}(x)\,u^{\alpha}\,.
\ee
With this change of variables from $(X,U)$ to $(R,x;v,u)$\,, the homogeneous and tangent condition \eqref{ht con} are solved by the dS generating functions
\be
	\Phi(R,x;v,u)= \left(\tfrac RL\right)^{\Delta_{h}}\,\varphi(x,u)\,,
\ee
and it is useful for later use to present the ambient-space derivatives expressed in the radial coordinates:
\ba
	&& \partial_{X^{M}}= \hat X_{\sst M}\,\partial_{R}
	+\tfrac{L^{2}}{R}\,\tfrac{\partial \hat X_{M}}{\partial x_{\mu}}
	\left[D_{\mu}+\tfrac1L\left( u\cdot e_{\mu}\,\partial_{v}
	-v\,\partial_u\!\cdot e_{\mu}\right)\,\right],
	\nn
	&& \partial_{U^{M}}= \hat X_{\sst M}\,\partial_{v}
	+L\, \tfrac{\partial \hat X_{M}}{\partial x_{\mu}}\,
	\partial_{u}\!\cdot e_{\mu}\,.
	\label{amb der}
\ea
Here $D_{\mu}$ is the dS covariant derivative defined in \eqref{cov}.

\paragraph{Gauge symmetries, Field equations and Actions}

So far the ambient-space tensor fields were not necessarily gauge fields. Now we want to address the case where the ambient-space fields admit linearized gauge symmetries of the form:
\be
	\delta^{\sst (0)}_{E}\,\Phi=U\cdot\partial_{X}\,E\,,
	\label{amb gt}
\ee
where $E(X,U)$ are generating functions of gauge parameters which
are homogeneous and tangent according to eq.~\eqref{ht con}.
From the expression of the gradient operator in radial coordinates:
\be
	U\cdot\partial_{X}=v\,\partial_{R}+
	\tfrac LR \left[ u\cdot e^{\mu}\,D_{\mu}+\tfrac1L\left(u^{2}\,\partial_{v}
	-v\,u\cdot\partial_{u}\right)\right],
\ee
one can see that choosing the degrees of homogeneity for gauge fields and gauge parameters as
\be
	\Phi(R,x;v,u)=
	\left(\tfrac RL\right)^{u\cdot\partial_{u}-2}
	\, \varphi(x,u)\,,
	\qquad
	E(R,x;v,u)=\left(\tfrac RL\right)^{u\cdot\partial_{u}}
	\, \varepsilon(x,u)\,,
	\label{reduction}
\ee
the ambient-space gauge transformations \eqref{amb gt}
induce the dS ones in eq.~\eqref{dS gt}.\footnote{It was shown in
\cite{Francia:2008hd} that different degrees of homogeneity
give the gauge transformations of partially massless fields.} Moreover, the TT constraints imposed
on the ambient-space gauge fields and gauge parameters are equivalent to the TT constraints on the dS ones.
Most importantly, the field equations of the ambient-space HS gauge fields also induce the field equations of the dS fields, given by
\be
	\partial_{X}^{2}\,\Phi=
	\left(\tfrac RL\right)^{u\cdot\partial_{u}-4}
	\left(D^{2}+\tfrac 1{L^{2}}\,\mu \right)
	\varphi\approx 0\,.
	\label{eom amb}
\ee
Hence, the dS action \eqref{AdS act} can be expressed in the ambient-space language as
\be
	S^{\sst (2)} = \frac1{2}
	\int \frac{d^{d+1}X}L\ \delta\Big(\tfrac{\sqrt{X^{2}}}L-1\Big)\,
	\Big[\ \delta_{a_{1}a_{2}}\,e^{\partial_{U_{1}}\!\cdot\,\partial_{U_{2}}}\,
	\Phi^{a_{1}}(X_{1},U_{1})\,\partial_{X_{2}}^{\,2}\,\Phi^{a_{2}}(X_{2},U_{2})
	\Big]_{\overset{X_{i}=X}{\sst U_{i}=0}}\,.\label{freeAdS}
\ee
This is because the ambient-space integral splits into the dS one together with an additional radial
integral as
\be
	\int \frac{d^{d+1}X}L
	=\int_{0}^{\infty} \tfrac{d R}L \left(\tfrac{R}{L}\right)^{d}\,\int_{\rm dS} d^{d}x\sqrt{-g}\,,
\ee
while the ambient-space contractions simply reduce to the dS contractions:
\be
	\partial_{U_{1}}\!\!\cdot \partial_{U_{2}}\,\big( \cdots\big)\,\big|_{X_{i}=X}
	=(\partial_{v_{1}}\partial_{v_{2}}+\partial_{u_{1}}\!\!\cdot \partial_{u_{2}}
	)\,\big(\cdots\big)\,\big|_{X_{i}=X}\,,
\ee
where of course $\partial_{v_{1}} \partial_{v_{2}}$ does not play a role
whenever the contraction acts on fields $\varphi$ that do not depend on $v$\,.

Finally, we want to mention that there are two ways of formulating the dS action in the ambient-space formalism: one with the insertion of the $\delta$-function and the other without it. In this paper we choose the former in order to avoid ambiguities which might arise from the formally diverging radial integrals.

\section{Cubic interactions of HS gauge fields in (A)dS backgrounds}
\label{sec: ads cubic}

In this section we construct the consistent cubic interactions of HS gauge fields in
(A)dS within the TT setting. The interaction vertices will be provided in the ambient-space formalism, and we shall then show how to obtain the (A)dS-intrinsic expressions by radially reducing the ambient-space ones.
We notice also that some key properties of the cubic vertices can be extracted
from the ambient-space forms without re-expressing them in terms of
intrinsic (A)dS quantities.

\subsection{Construction of consistent cubic interactions}

Our construction of the consistent cubic interactions in (A)dS
is based on the ambient-space formalism,
and let us notice that by considering all possible forms of ambient-space cubic vertices,
one does not loose any generality since any dS vertex can be written as an ambient-space one making use of the following projector\footnote{
See e.g. \cite{Bekaert:2010hk} for more details.}:
\be
	P^{\sst MN}=\frac{X^{\sst M}\,X^{\sst N}-\eta^{\sst MN}\,X^{2} }{X^{2}}\,.
\ee
However, as is suggested by the projector itself,
the possible ambient-space cubic vertices are a bit more general
than the flat-space ones.
This can be also understood as a result of the break down of translational symmetries at the level of the ambient space induced by the $\d$-function insertion. Therefore, the most general expression of
the cubic interactions acquires one more argument, $X^{\sst M}$\,,
compared to the flat-space case \eqref{f ansz}:
\ba
	S^{\sst (3)} \eq
	\frac1{3!}\int \frac{d^{d+1}X}L\
	\delta\Big(\tfrac{\sqrt{X^{2}}}L-1\Big)\
	C_{a_{1}a_{2}a_{3}}(\,X\, ;\,
	\partial_{X_1}\,,\,\partial_{X_{2}}\,,\,
	\partial_{X_3}\,;\,\partial_{U_{1}}\,,\,
	\partial_{U_2}\,,\,\partial_{U_{3}}\,)\times \nn
	&& \hspace{120pt}\times\,
	\Phi^{a_{1}}(X_{\sst 1},U_{\sst 1})\ \Phi^{a_{2}}(X_{\sst 2},U_{\sst 2})\
	\Phi^{a_{3}}(X_{\sst 3},U_{\sst 3})\,\Big|_{\overset{X_{i}=X}{\sst U_{i}=0}}\,.
\ea
Since the arguments of $C_{a_{1}a_{2}a_{3}}$ do not commute,
we decide to choose the ordering where all $X^{\sst M}$'s are placed on the left side of
$\partial_{X_{i}}$'s and $\partial_{\,U_{j}}$'s\,.

As in the flat-space case, we can simplify the ansatz making use of
all possible Lorentz invariants. However, compared to the flat case
we have a few more quantities
$X^{2}$\,, $X\cdot\partial_{X_{i}}$ and $X\cdot \partial_{\,U_{i}}$ and
they should be placed on the left side of other Lorentz invariants in our conventions.
In the following, we show that the dependence on these additional Lorentz invariant objects in the ansatz
can be in fact neglected.
\begin{itemize}
\item
First, $X^{2}$ simply becomes $L^{2}$ after the radial integration,
and can be absorbed into the definition of $C_{a_{1}a_{2}a_{3}}$\,.
\item
Second, $X\cdot \partial_{X_{i}}$ is equivalent to $X_{i}\cdot\partial_{X_{i}}$ which essentially counts the number of $X_{i}$'s and so can be absorbed into $C_{a_{1}a_{2}a_{3}}$ as well.
\item
Finally, $X\cdot \partial_{U_{i}}$ is equivalent to $X_{i}\cdot\partial_{U_{i}}$ which is nothing but the tangent condition \eqref{ht con}. Hence, when it acts directly on fields it vanishes while
acting on the derivatives produces
\ba
	&& (X\cdot\partial_{U})\ \partial_{X^{\sst M_{1}}}\,\cdots\,\partial_{X^{\sst M_{n}}}\,\Phi \nn
	&& \quad =-
	\sum_{m=1}^{n}\,\partial_{X^{\sst M_{1}}}\,\cdots\,\partial_{X^{\sst M_{m-1}}}\,
	\partial_{U^{\sst M_{m}}}\,
	\partial_{X^{\sst M_{m+1}}}\,\cdots\,\partial_{X^{\sst M_{m-1}}}\,\Phi\,,
\ea
so that $X\cdot \partial_{U_{i}}$ is equivalent to a linear combination of the other Lorentz invariants.
\end{itemize}

After excluding these $X^{\sst M}$-depending Lorentz invariants, one ends up with the same ansatz \eqref{f ansz} which was the starting point of the flat-space construction. However, there is one other difference: total-derivative terms do not simply vanish due to the presence of the $\delta$-function insertion, and they do contribute to the action.
To properly analyze the role of the total derivatives, the latter is denoted by
\mt{\partial_{X^{\sst M}}=\partial_{X_{1}^{\sst M}}+\partial_{X_{2}^{\sst M}}+\partial_{X_{3}^{\sst M}}} and the other non-total derivatives by \mt{\partial_{X_{ij}^{\sst M}}:=\partial_{X^{\sst M}_{i}}-\partial_{X^{\sst M}_{j}}}\,.
Then the ansatz for the cubic interactions is given by
\be
	C_{a_{1}a_{2}a_{3}}(\,\partial_{X}\, ;\,
	\partial_{X_{12}}\,,\,\partial_{X_{23}}\,,\,
	\partial_{X_{31}}\,;\,\partial_{U_{1}}\,,\,
	\partial_{U_2}\,,\,\partial_{U_{3}})\,,
	\label{C dS}
\ee
and this can be expanded in the powers of the total derivatives.
By noticing that the total derivatives can be integrated by parts as
\be
	\delta^{\sst (n)}\Big(\tfrac{\sqrt{X^{2}}}L-1\Big)\
	\partial_{X^{\sst M}}
	=-\frac1{L^{2}}\,\delta^{\sst(n+1)}\Big(\tfrac{\sqrt{X^{2}}}L-1\Big)\ X^{\sst M}\,,
	\label{int bp}
\ee
where $\delta^{\sst (n)}(x):=\left(\frac{d}{dx}\right)^{n}\delta(x)$\,,
and that the $X^{\sst M}$-dependence in the vertex can be eliminated
as we argued before,
the general ansatz \eqref{C dS} for the cubic vertices can be recast
in the form
\be
	\delta\Big(\tfrac{\sqrt{X^{2}}}L-1\Big)\
	C_{a_{1}a_{2}a_{3}}(\,\partial_{X}\, ;\,
	\partial_{X_{ij}},\partial_{U_{i}})
	=\sum_{n=0}^{\infty}\,\frac1{L^{n}}\, \delta^{\sst (n)}\Big(\tfrac{\sqrt{X^{2}}}L-1\Big)\,
	C_{a_{1}a_{2}a_{3}}^{{\sst (n)}}(\,\partial_{X_{ij}},\partial_{U_{i}})\,.
	\label{delta exp}
\ee
Similarly to the flat-space construction, $C^{\sst (n)}_{a_{1}a_{2}a_{3}}$ can be
written as a function of the Lorentz invariants $\cY_{i}$'s and $\cZ_{i}$'s, which are now the ambient-space
analogues of \eqref{Y&Z}:
\be
	C_{a_{1}a_{2}a_{3}}^{{\sst (n)}}(\,\partial_{X_{ij}},\partial_{U_{i}})
	=
	C_{a_{1}a_{2}a_{3}}^{{\sst (n)}}(\,\cY_{i}\,,\,\cZ_{i}\,)\,.
\ee
Making use of eq.~\eqref{delta exp}, the gauge consistency condition reads
\be
	\sum_{n=0}^{\infty}\,\frac1{L^{n}}\, \delta^{\sst (n)}\Big(\tfrac{\sqrt{X^{2}}}L-1\Big)\,
	\big[\, C_{a_{1}a_{2}a_{3}}^{{\sst (n)}}(\,\cY_{i}\,,\,\cZ_{i}\,)\,,
	\,U_{\sst j}\!\cdot\partial_{X_{\sst j}}\,\big] \approx 0\,,
	\label{GC dS}
\ee
where $\approx$ means equivalence modulo the free field equations
$\partial_{X_{i}}^{2}\approx 0$\,.
Moreover, the commutator in eq.~\eqref{GC dS} can be simplified  making use of the identities:
\be
	\big[\, \cY_{i}\,,\,U_{\sst j}\!\cdot\partial_{X_{j}}\,\big] \approx
	\tfrac12\,\delta_{ij}\,\epsilon_{ik\ell}\,
	\partial_{X} \!\cdot \partial_{X_{k\ell}}\,, \quad
	\big[\,\cZ_{i}\,,\,U_{\sst j}\!\cdot\partial_{X_{j}}\,\big] = -\tfrac12\,\epsilon_{ijk}\,\cY_{k}+
	\tfrac12\,|\epsilon_{ijk}|\,\partial_{X}\!\cdot\partial_{U_{k}}\,,
\ee
where the total derivatives terms play a role compared to the flat-space case \eqref{YZ id}
since they contribute to the $\delta^{\sst (n+1)}$-order after an integration by parts.
Taking into account the other commutator identities:
\ba
	&&
	\big[\,X_{1}\!\cdot\partial_{U_{1}}\,,\,F(\,\cY_{i}\,,\,\cZ_{i}\,)\,\big]=
	\left(\cZ_{3}\,\partial_{\cY_{\sst 2}}-\cZ_{2}\,\partial_{\cY_{\sst 3}}
	\right) F(\,\cY_{i}\,,\,\cZ_{i}\,)\,,\nn
	&&
	\big[\,X_{1}\!\cdot\partial_{X_{1}}\,,\,F(\,\cY_{i}\,,\,\cZ_{i}\,)\,\big]=
	-\tfrac12\left[
	\left(\cY_{2}-\partial_{X}\!\cdot\partial_{U_{\sst 2}}\right)
	\,\partial_{\cY_{\sst 2}}+
	\left(\cY_{3}
	+\partial_{X}\!\cdot\partial_{U_{\sst 3}}\right)
	\,\partial_{\cY_{\sst 3}}
	\right] F(\,\cY_{i}\,,\,\cZ_{i}\,)\,,\nn
	&&
	\big[\,F(\,\cY_{i}\,,\,\cZ_{i}\,)\,,\,U_{\sst 1}\!\cdot\partial_{U_{1}}\,\big]=
	\left(\cY_{1}\,\partial_{\cY_{\sst1}}+
	\cZ_{2}\,\partial_{\cZ_{\sst 2}}+\cZ_{3}\,\partial_{\cZ_{\sst 3}}
	\right) F(\,\cY_{i}\,,\,\cZ_{i}\,)\,,
\ea
we finally get the following differential equation and its cyclic permutations on
$(1)(2)(3)$\,:
\ba
	&& \left(\cY_{1}\,\partial_{\cZ_{\sst 2}}-\cY_{2}\,\partial_{\cZ_{\sst 1}}\right) C^{\sst (n)}_{a_{1}a_{2}a_{3}} \nn
	&&
	+\,\tfrac1{L} \left[
	3 \left(\cY_{1}\,\partial_{\cY_{\sst 1}}-\cY_{2}\,\partial_{\cY_{\sst 2}}\right) \partial_{\cY_{\sst 3}}
	-2\,\cZ_{3}\left(\partial_{\cY_{\sst 1}}\,\partial_{\cZ_{\sst 1}}-
	\partial_{\cY_{\sst 2}}\,\partial_{\cZ_{\sst 2}}\right) \right] C^{\sst (n-1)}_{a_{1}a_{2}a_{3}} \nn
	&& -\,\tfrac1{L^{2}}\,3\left(\cZ_{1}\,\partial_{\cY_{\sst 2}}-\cZ_{2}\,\partial_{\cY_{\sst 1}}\right)\partial_{\cY_{\sst 3}}^{\ 2}
	\,C^{\sst (n-2)}_{a_{1}a_{2}a_{3}}
	=0\,.
	\label{dfe dS}
\ea
Here $C^{\sst (n)}_{a_{1}a_{2}a_{3}}$ vanishes for a negative $n$\,,
and in the $L\to \infty$ limit
these equations reduce to the flat-space condition \eqref{dfe c}.

Finally, the consistent parity-invariant cubic interactions in $(A)dS_{d}$ for $d\ge4$
can be obtained as solutions of the differential recurrence equations \eqref{dfe dS},
which can be solved from \mt{n=0} iteratively.
For \mt{n=0} the corresponding equation coincides with the flat-space one \eqref{dfe c}, and its solutions are  given again
by arbitrary functions $\cK_{a_{1}a_{2}a_{3}}$ as
\be
	C^{\sst (0)}_{a_{1}a_{2}a_{3}}(\,\cY_{i}\,,\,\cZ_{i}\,)
	= \cK_{a_{1}a_{2}a_{3}}(\,\cY_{i}\,,\,\cG\,)\,,
	\label{sol dS 0}
\ee
where $\cG:=\cY_{i}\,\cZ_{i}$\,.
On the contrary, for each \mt{n\ge1} it is an inhomogeneous differential equation
whose solutions are fixed up to a solution of the corresponding homogeneous equation.
Each of these homogeneous solutions $\cK'_{a_{1}a_{2}a_{3}}(\cY_{i},\cG)$ induces a tail of particular solutions for higher orders, and provides
additional solutions of eqs.~\eqref{dfe dS}:
\be
	C_{a_{1}a_{2}a_{3}}=
	\frac1{L^{n}}\,\delta^{\sst (n)}\Big(\tfrac{\sqrt{X^{2}}}L-1\Big)\,\cK'_{a_{1}a_{2}a_{3}}
	+\sum_{m=n+1}^{\infty} \frac1{L^{m}}\,
	\delta^{\sst (m)}\Big(\tfrac{\sqrt{X^{2}}}L-1\Big)\,
	C^{\sst (m)}_{a_{1}a_{2}a_{3}}\,.
	\label{C ser}
\ee
This observation seems to imply the presence of ambiguities in the (A)dS interactions,
but these additional solutions are in fact redundancies.
This is because, after the radial integration,
different $\delta^{\sst (n)}$'s give just different spin-dependent constant factors.
Therefore, any solution of the type \eqref{C ser} can be re-expressed in the form:
\be
	C_{a_{1}a_{2}a_{3}} =
	\delta\Big(\tfrac{\sqrt{X^{2}}}L-1\Big)\,\tilde\cK'_{a_{1}a_{2}a_{3}}
	+\sum_{n=1}^{\infty}\,\frac1{L^{n}}\,\delta^{\sst (n)}\Big(\tfrac{\sqrt{X^{2}}}L-1\Big)\,
	\tilde C^{\sst (n)}_{a_{1}a_{2}a_{3}}\,,
\ee
where $\tilde \cK'_{a_{1}a_{2}a_{3}}$ and $\tilde C^{\sst (n)}_{a_{1}a_{2}a_{3}}$
are proportional to $\cK'_{a_{1}a_{2}a_{3}}$ and $C^{\sst (n)}_{a_{1}a_{2}a_{3}}$
with some spin-dependent factors.
To reiterate, the aforementioned ambiguity can be recast into a redefinition of the original
\mt{C^{\sst (0)}_{a_{1}a_{2}a_{3}}=\cK_{a_{1}a_{2}a_{3}}}\,.
Hence, a consistent cubic interaction is univocally determined from the choice of
 $C^{\sst (0)}_{a_{1}a_{2}a_{3}}$\,.

\subsection{General solutions for consistent cubic interactions}

So far, we have shown that the consistent cubic interactions in (A)dS
can be obtained solving the differential equations \eqref{dfe dS}.
The $\delta^{\sst(0)}$-order solution was already obtained in terms of an arbitrary function
$\cK_{a_{1}a_{2}a_{3}}$\,, and what is left is to determine a particular solution for
the higher order parts $C^{\sst (n\ge 1)}_{a_{1}a_{2}a_{3}}$\,,
keeping in mind that the ambiguities in the latter are redundancies.

In the following, we construct at once the full cubic vertex
$C_{a_{1}a_{2}a_{3}}(\,\partial_{X}\, ;\,\partial_{X_{ij}},\partial_{U_{i}})$\,,
comprising the full higher order tail of the $C^{\sst (n)}_{a_{1}a_{2}a_{3}}$'s,
by making use of the following ansatz:
\be
	C_{a_{1}a_{2}a_{3}}(\,\partial_{X}\, ;\,
	\partial_{X_{ij}},\partial_{U_{i}})=
	\cK_{a_{1}a_{2}a_{3}}\big(\, \cY^{\sst \Lambda}_{i}\,,\,\cG^{\sst \Lambda}\,\big)\,,
	\label{amb K}
\ee
where $\cY^{\sst \Lambda}_{i}$'s and $\cG^{\Lambda}$ are given by
\ba
	\cY^{\sst \Lambda}_{i} \eq \cY_{i}+\alpha_{i}\,\partial_{U_{i}}\!\!\cdot\partial_{X}\,, \nn
 	\cG^{\sst \Lambda} \eq
	(\cY_{1} + \beta_{1}\,\partial_{U_{1}}\!\!\cdot\partial_{X})\,\cZ_{1} +
	(\cY_{2} + \beta_{2}\,\partial_{U_{2}}\!\!\cdot\partial_{X})\,\cZ_{2} +
	(\cY_{3} + \beta_{3}\,\partial_{U_{3}}\!\!\cdot\partial_{X})\,\cZ_{3}\,.
	\label{amb G}
\ea
Notice first that this ansatz is a highly restricted one, with only six constants,
compared to the general setting with an arbitrary number of $C^{\sst (n)}_{a_{1}a_{2}a_{3}}$'s.
Nonetheless, the motivation is straightforward: we attempt to keep
the form of the generating function $\cK_{a_{1}a_{2}a_{3}}$ fixed.
Notice as well that, although \eqref{amb G} contains explicitly
total derivatives, the highest-derivative part of the vertices built from
\eqref{amb G} do not, ensuring its non-triviality.

In order to examine the ansatz \eqref{amb K},
we compute the gauge variation of the latter
exercising some care in treating total derivatives.
We first provide our solutions, leaving
the detailed computation for the last part of this subsection.
 Requiring that the gauge variation vanish modulo $\partial_{X_{i}}^{2}\approx 0$\,, one ends up with the conditions %
\ba
	&& (\alpha_{1}+1)(\alpha_{2}-1)+4=0\,, \label{B1} \\
	&& (\alpha_{1}+1)(\beta_{2}+1)+(\alpha_{1}-1)(\beta_{3}-1)=0\,,\label{B2} \\
	&& (\beta_{1}+1)(\beta_{2}+1)+(\beta_{3}-1)(\beta_{1}+\beta_{2})=0\,, \label{B3}
\ea
on the constants $\alpha_{i}$ and $\beta_{i}$ appearing in the ansatz \eqref{amb K},
together with their cyclic permutations on the subscripts $i$ of $\alpha_{i}$ and $\beta_{i}$\,. If a solution for these equations exists,
eq.~\eqref{amb K} provides the consistent cubic interactions. Actually, eqs.~(\ref{B1}\,-\,\ref{B3}) admit solutions parameterized by two constants $\alpha$ and $\beta$\,:
\ba
	 \alpha_{1}=\alpha\,,\qquad
	&\alpha_{2}=\tfrac{\alpha-3}{\alpha+1}\,,\qquad
	&\alpha_{3}=-\tfrac{\alpha+3}{\alpha-1}\,, \nn
	 \beta_{1}=\beta\,,\qquad
	&\beta_{2}=2\,\tfrac{\alpha-\beta}{\alpha+1}-1\,,\qquad
	&\beta_{3}=-2\,\tfrac{\alpha-\beta}{\alpha-1}+1\,.
	\label{sol cnst}
\ea
%
As we have argued in the previous section, different choices of $\alpha$ and $\beta$
(which give different solutions for $C^{\sst (n\ge1)}_{a_{1}a_{2}a_{3}}$) can be reabsorbed into the definition
of the functions $\cK_{a_{1}a_{2}a_{3}}$
so that one can work with a particular choice of $\alpha$ and $\beta$ without loss of generality.

Finally, the general solution for the (A)dS cubic-interaction problem is given by an arbitrary function
$\cK_{a_{1}a_{2}a_{3}}$ \eqref{amb K} together with eq.~\eqref{sol cnst}\,, and when the latter is expanded for different spins, we obtain the expression in eq.~\eqref{sol ads} with the coupling constants $g^{s_{1}s_{2}s_{3},n}_{a_{1}a_{2}a_{3}}$\,.
One can easily verify that when \mt{s_{1}=s_{2}=0} our results coincide
with the interaction vertices constructed in \cite{Bekaert:2010hk}. The coupling constants have the same mass-dimensions \eqref{cc md}
and the same permutation symmetries \eqref{cc cyc} as the flat-space ones, while each vertex is now not homogeneous in the number of the (A)dS-covariant derivatives since the ambient-space derivative \eqref{amb der} is not.
However, the maximum number of derivatives of the vertex associated with $g^{s_{1}s_{2}s_{3},n}_{a_{1}a_{2}a_{3}}$ can be easily identified as \mt{s_{1}+s_{2}+s_{3}-2n}\,. In the next section we shall see in more detail how this non-homogeneity is related to
the inverse power expansion in the cosmological constant exhibited by the Fradkin-Vasiliev vertices.

\subsubsection*{Proof at  the $\delta^{\sst (1)}$ level}

Here we show that  the gauge invariance of the ansatz \eqref{amb K} is equivalent to the conditions of eqs.~(\ref{B1}\,-\,\ref{B3})\,. Since all vertices of different spins are independent, we consider without loss of generality the case where $\cK^{\Lambda}_{a_{1}a_{2}a_{3}}$ is given by an exponential function:
\be
	\cK^{\Lambda}_{a_{1}a_{2}a_{3}}=k_{a_{1}a_{2}a_{3}}\,e^{L\,\cV}\,,
\ee
where $L$ is again the radius of dS and $\cV$ is the sum of the arguments of $\cK_{a_{1}a_{2}a_{3}}$ in
eq.~\eqref{amb K}\,:
\ba
	\cV \edf \partial_{U_{1}}\!\!\cdot(\partial_{X_{23}}\!+\alpha_{1}\,\partial_{X})+
	\partial_{U_{2}}\!\!\cdot(\partial_{X_{31}}\!+\alpha_{2}\,\partial_{X})+
	\partial_{U_{3}}\!\!\cdot(\partial_{X_{12}}\!+\alpha_{3}\,\partial_{X}) \nn
	&&+\, \partial_{U_{2}}\!\!\cdot\partial_{U_{3}}\,
	\partial_{U_{1}}\!\!\cdot(\partial_{X_{23}}\!+\beta_{1}\,\partial_{X})
	+\partial_{U_{3}}\!\!\cdot\partial_{U_{1}}\,\partial_{U_{2}}\!\!\cdot
	(\partial_{X_{31}}\!+\beta_{2}\,\partial_{X})\nn
	&&+\,\partial_{U_{1}}\!\!\cdot\partial_{U_{2}}\,\partial_{U_{3}}\!\!\cdot
	(\partial_{X_{12}}\!+\beta_{3}\,\partial_{X})\,.
\ea
It is now convenient to use the following compact notation for the cubic action:
\be
	S^{\sst (3)}=\frac1{3!}\int \ \delta\
	k\ e^{L\,\cV}\ \Phi_{1}\,\Phi_{2}\,\Phi_{3}\,\big|\,,
\ee
where we use as a shorthand notation $k$ and $\Phi_{i}$ in place of
$k_{a_{1}a_{2}a_{3}}$ and $\Phi^{a_{i}}(X_{i}, U_{i})$\,, and $|$ at the end of equation denotes
the evaluation $X_{i}=X$ and $U_{i}=0$\,. Performing the gauge variation with respect to $\Phi_{1}$\,, one then gets
\be
	\delta^{\sst (0)}_{E_{1}}\,S^{\sst (3)} = \frac1{3!}\int \ \delta\
	k \big[\,\cV\,,\,U_{1}\!\cdot\partial_{X_{1}}\big]\ e^{L\,\cV}
	\ E_{1}\,\Phi_{2}\,\Phi_{3}\,\big|\,,
\ee	
where the commutator $[\,\cV\,,\,U_{1}\!\cdot\partial_{X_{1}} ]$ is given by the total derivative terms:
\ba
	&& \big[\,\cV\,,\,U_{1}\!\cdot\partial_{X_{1}}\big] \approx
	\partial_{X}\!\cdot\partial_{X_{23}}\,(1+\partial_{U_{2}}\!\!\cdot\partial_{U_{3}})
	+\partial_{X}\!\cdot\partial_{X_{1}}\,(\alpha_{1}+\beta_{1}\,\partial_{U_{2}}\!\!\cdot\partial_{U_{3}})\nn
	&& \hspace{20pt} +\,
	\partial_{X}\!\cdot\partial_{U_{2}}\,\big(\tfrac12\,\partial_{X_{12}}\!\!\cdot\partial_{U_{3}}
	+\beta_{2}\,\partial_{X_{1}}\!\!\cdot\partial_{U_{3}}\big)
	+\partial_{X}\!\cdot\partial_{U_{3}}\,\big(\tfrac12\,\partial_{X_{31}}\!\!\cdot\partial_{U_{2}}
	+\beta_{3}\,\partial_{X_{1}}\!\!\cdot\partial_{U_{2}}\big)\,. \qquad
\ea
After integrations by parts, one ends up with terms proportional to $X_{i}\!\cdot\partial_{X_{i}}$ and $X_{i}\!\cdot\partial_{U_{i}}$ which are exactly the operators appearing in the homogeneous and tangent conditions \eqref{ht con}. To make these operators act directly on the fields, one must move them to the right of $e^{L\,\cV}$ using the commutators $[\,X_{i}\!\cdot\partial_{X_{i}}\,,\cV\,]$ and $[\,X_{i}\!\cdot\partial_{U_{i}}\,,\cV\,]$\,. As a result one gets
\ba
	\delta^{\sst (0)}_{E_{1}}\,S^{\sst (3)} \approx -\frac1{3!}\int \ \delta^{\sst (1)}\
	k\ e^{L\,\cV}\ \cA\ E_{1}\,\Phi_{2}\,\Phi_{3}\,\big|\,,
\ea
with
\ba
	&& \cA =
	\big(1+\partial_{U_{2}}\!\!\cdot\partial_{U_{3}}\big)\,
	\big[\,\tfrac1L(X_{2}\!\cdot\partial_{X_{2}}-X_{3}\!\cdot\partial_{X_{3}})
	-(\alpha_{1}\,\partial_{X_{23}}+\partial_{X})\!\cdot\partial_{U_{1}}
	+(\alpha_{2}+1)\,\partial_{X_{3}}\!\!\cdot\partial_{U_{2}} \nn
	&& \hspace{100pt} -\,(\alpha_{3}-1)\,\partial_{X_{2}}\!\!\cdot\partial_{U_{3}}
	-(\beta_{1}\,\partial_{X_{23}}+\partial_{X})\!\cdot\partial_{U_{1}}\,\partial_{U_{2}}\!\!\cdot\partial_{U_{3}} \nn
	&&\hspace{100pt} +\,
	(\beta_{2}+1)\,\partial_{X_{3}}\!\!\cdot\partial_{U_{2}} \,\partial_{U_{3}}\!\!\cdot\partial_{U_{1}}
	-(\beta_{3}-1)\,\partial_{X_{2}}\!\!\cdot\partial_{U_{3}}\,\partial_{U_{1}}\!\!\cdot\partial_{U_{2}}\, \big]\nn
	&&\quad +\,\big(\alpha_{1}+\beta_{1}\,\partial_{U_{2}}\!\!\cdot\partial_{U_{3}}\big)\,
	\big[\,\tfrac1L\,X_{1}\!\cdot\partial_{X_{1}}-(\alpha_{2}-1)\,\partial_{X_{1}}\!\!\cdot\partial_{U_{2}}
	-(\alpha_{3}+1)\,\partial_{X_{1}}\!\!\cdot\partial_{U_{3}}
	\nn
	&& \hspace{115pt}
	-\,(\beta_{2}-1)\,\partial_{X_{1}}\!\!\cdot\partial_{U_{2}} \,\partial_{U_{3}}\!\!\cdot\partial_{U_{1}}
	-(\beta_{3}+1)\,\partial_{X_{1}}\!\!\cdot\partial_{U_{3}}\,\partial_{U_{1}}\!\!\cdot\partial_{U_{2}}\, \big] \\
	&&\quad +\,\big(\tfrac12\,\partial_{X_{12}}\!\!\cdot\partial_{U_{3}}
	+\beta_{2}\,\partial_{X_{1}}\!\!\cdot\partial_{U_{3}}\big)\,
	\big[\,\tfrac1L\,X_{2}\!\cdot\partial_{U_{2}}
	-(\alpha_{3}-1)\,\partial_{U_{2}}\!\!\cdot\partial_{U_{3}}
	-(\alpha_{1}+1)\,\partial_{U_{1}}\!\!\cdot\partial_{U_{2}}
	\nn
	&& \hspace{155pt}
	-\,(\beta_{3}-1)\,\partial_{U_{2}}\!\!\cdot\partial_{U_{3}} \,\partial_{U_{1}}\!\!\cdot\partial_{U_{2}}
	-(\beta_{1}+1)\,\partial_{U_{1}}\!\!\cdot\partial_{U_{2}}\,\partial_{U_{2}}\!\!\cdot\partial_{U_{3}}\, \big]
	 \nn
	&&\quad +\,\big(\tfrac12\,\partial_{X_{31}}\!\!\cdot\partial_{U_{2}}
	+\beta_{3}\,\partial_{X_{1}}\!\!\cdot\partial_{U_{2}}\big)\,
	\big[\,\tfrac1L\,X_{3}\!\cdot\partial_{U_{3}}
	-(\alpha_{1}-1)\,\partial_{U_{3}}\!\!\cdot\partial_{U_{1}}
	-(\alpha_{2}+1)\,\partial_{U_{2}}\!\!\cdot\partial_{U_{3}}
	\nn
	&& \hspace{155pt}
	-\,(\beta_{1}-1)\,\partial_{U_{3}}\!\!\cdot\partial_{U_{1}} \,\partial_{U_{2}}\!\!\cdot\partial_{U_{3}}
	-(\beta_{2}+1)\,\partial_{U_{2}}\!\!\cdot\partial_{U_{3}}\,\partial_{U_{3}}\!\!\cdot\partial_{U_{1}}\, \big]\,.
	\nonumber
\ea
The resulting term $\cA$ contains $X_{i}\!\cdot \partial_{X_{i}}$ and $X_{i}\!\cdot \partial_{U_{i}}$ as well as other terms coming from
the commutation relations. Here $X_{i}\!\cdot \partial_{U_{i}}$ vanishes by the tangent condition, and the homogenous condition replaces
\mt{X_{1}\!\cdot\partial_{X_{1}}} and \mt{X_{2}\!\cdot\partial_{X_{2}}-X_{3}\!\cdot\partial_{X_{3}}}
respectively, with \mt{U_{1}\!\cdot\partial_{U_{1}}} and \mt{U_{2}\!\cdot\partial_{U_{2}}-U_{3}\!\cdot\partial_{U_{3}}}\,. Since the last two depend on $U_{i}$\,, pushing them to the left of $e^{L\,\cV}$\,, they vanish when evaluated at $U_{i}=0$ and the only remaining
contributions are the commutators. Collecting all the resulting terms one finally ends up with
\ba
	\delta^{\sst (0)}_{E_{1}}\,S^{\sst (3)} \approx -\frac1{3!}\int\ \delta^{\sst (1)}\
	k\ (\,\cB+\cC\,)\ e^{L\,\cV}\ E_{1}\,\Phi_{2}\,\Phi_{3}\,\big|\,,
	\label{gv end}
\ea
where we have separated terms into the non-total-derivative part $\cB$ (which involves only $\partial_{X_{ij}^{\sst M}}$ but not $\partial_{X^{\sst M}}$) and the total-derivative part $\cC$ (which contains $\partial_{X^{\sst M}}$). If the gauge variation $\delta_{1}\,S^{\sst (3)}$ vanishes, $\cB$ should vanish as well since there is no way to compensate it. In order to simplify the discussion one can split $\cB$ as $\cB=\cB_{1}+\cB_{2}+\cB_{3}$, where the $\cB_{n}$'s are of order $n$ in the Lorentz invariants and are given respectively by
\ba
	\cB_{1}\eq \tfrac12\,\big[ (\alpha_{1}+1)(\alpha_{2}-1)+4\big]\,\partial_{X_{31}}\!\!\cdot\partial_{U_{2}}
	-\tfrac12\,\big[ (\alpha_{3}+1)(\alpha_{1}-1)+4\big]\,\partial_{X_{12}}\!\!\cdot\partial_{U_{3}}\,,
	\label {cB1} \\
	\cB_{2} \eq \tfrac12\,\big[(\alpha_{1}+1)(\beta_{2}+1)+(\alpha_{1}-1)(\beta_{3}-1)\big]
	\big(\partial_{X_{31}}\!\!\cdot\partial_{U_{2}}\,\partial_{U_{3}}\!\!\cdot\partial_{U_{1}}
	-\partial_{X_{12}}\!\!\cdot\partial_{U_{3}}\,\partial_{U_{1}}\!\!\cdot\partial_{U_{2}}\big) \nn
	&&+\,\tfrac12\,\big[(\alpha_{2}-1)(\beta_{1}+1)+(\alpha_{2}+1)(\beta_{3}-1)+4\big]\,
	\partial_{X_{31}}\!\!\cdot\partial_{U_{2}}\,\partial_{U_{2}}\!\!\cdot\partial_{U_{3}} \nn
	&&-\,\tfrac12\,\big[(\alpha_{3}-1)(\beta_{2}+1)+(\alpha_{3}+1)(\beta_{1}-1)+4\big]\,
	\partial_{X_{12}}\!\!\cdot\partial_{U_{3}}\,\partial_{U_{1}}\!\!\cdot\partial_{U_{2}}\,,
	\label{cB2} \\
	\cB_{3}\eq \partial_{U_{2}}\!\!\cdot\partial_{U_{3}}\,\Big\{\,
	\tfrac12\,\big[(\beta_{1}+1)(\beta_{2}+1)+(\beta_{3}-1)(\beta_{1}+\beta_{2})\big]\,
	\partial_{X_{31}}\!\!\cdot\partial_{U_{2}}\,\partial_{U_{2}}\!\!\cdot\partial_{U_{3}} \nn
	&& \hspace{47pt}-\,\tfrac12\,\big[(\beta_{3}+1)(\beta_{1}+1)+(\beta_{2}-1)(\beta_{3}+\beta_{1})\big]\,
	\partial_{X_{12}}\!\!\cdot\partial_{U_{3}}\,\partial_{U_{1}}\!\!\cdot\partial_{U_{2}}\,\Big\}\,.
	\label{cB3}
\ea
Since $\cB_{n}$'s are independent, each $\cB_{n}$ should vanish separately. Moreover, since we have considered so far the gauge consistency only with respect to $\delta^{\sst (0)}\,\Phi^{a_{1}}(X_{1},U_{1})$\, we have still to take into account the gauge invariance with respect to $\delta^{\sst (0)}\,\Phi^{a_{2}}(X_{2},U_{2})$ and $\delta^{\sst (0)}\,\Phi^{a_{3}}(X_{3},U_{3})$\,. These give the same conditions just recovered here for $\delta^{\sst (0)}\,\Phi^{a_{1}}(X_{1},U_{1})$ but with cyclic permutations on the subscripts $i$ of $\alpha_{i}$ and $\beta_{i}$\,. Finally, the equations $\cB_{1}=0$\,,
$\cB_{2}=0$  and $\cB_{3}=0$ give respectively
the conditions \eqref{B1}, \eqref{B2} and \eqref{B3} on the constants $\alpha_{i}$ and $\beta_{i}$\,.

To complete the proof, one should also compute the total-derivative part $\cC$ in \eqref{gv end} and verify whether it imposes additional constraints on the
$\alpha_{i}$ and $\beta_{i}$\,. Actually, $\cC$ is vanishing with the conditions (\ref{B1}\,-\,\ref{B3}), and hence
the latter equations are sufficient. However, this cannot be seen simply at the present level $\delta^{\sst (1)}$\,, but needs to be carefully analyzed at the next level $\delta^{\sst (2)}$\,. The details of the proof
can be found in the Appendix \ref{sec:C=0}\,.

\subsection{Reduction to (A)dS-intrinsic expressions}

The cubic vertices \eqref{sol ads} constructed in the ambient-space formalism are given in terms of the Lorentz invariants \mt{\partial_{U_{i}}\!\!\cdot\partial_{X_{j}}} and \mt{\partial_{U_{i}}\!\!\cdot\partial_{U_{j}}}\,, and the expressions are compact but implicit with respect to (A)dS.
The explicit expressions in terms of (A)dS-intrinsic quantities can be obtained making use of the radial reduction formulas \eqref{amb der} and \eqref{reduction}. A convenient way for the reduction is to express the Lorentz invariants in terms of the following (A)dS-intrinsic bi-local quantities:
\ba
	Z(x_{i},x_{j}) \edf \hat X_{\sst M}(x_{i})\,\hat X^{\sst M}(x_{j})\,,\\
	H_{\mu}(x_{i},x_{j}) \edf L\,\frac{\partial \hat X_{\sst M}(x_{i})}{\partial x_{i}^{\mu}}\,\hat X^{\sst M}(x_{j})\,,\\
	G_{\mu\nu}(x_{i},x_{j}) \edf L^{2}\,
	\frac{\partial \hat X_{\sst M}(x_{i})}{\partial x_{i}^{\mu}}
	\frac{\partial \hat X^{\sst M}(x_{j})}{\partial x_{j}^{\nu}}\,,
	\label{bilocal}
\ea
whose coincident-point limits are given by
\be
	Z(x,x)=1\,, \qquad H^{\mu}(x,x)=0\,, \qquad G^{\mu\nu}(x,x)=g^{\mu\nu}(x)\,.
	\label{prop bilocal}
\ee
Here the indices of the bi-local quantities are raised or lowered with the local metric tensor.
With these conventions, the ambient-space Lorentz invariant operators can be written as
\ba
	\partial_{U_{i}}\!\cdot\partial_{X_{j}}\eq
	\Big[\partial_{v_{i}}\,Z(x_{i},x_{j})+\partial_{u_{i}^{\mu}}\,H^{\mu}(x_{i},x_{j})\Big]\,
	\partial_{R_{j}} \nn
	&& +\,
	\Big[\partial_{v_{i}}\,H^{\nu}(x_{j},x_{i})+\partial_{u_{i}^{\mu}}\,G^{\mu\nu}(x_{i},x_{j})\Big]
	\Big[D_{j\,\nu}+\tfrac1L\,(u_{j\,\nu}\,\partial_{v_{j}}-v_{j}\,\partial_{u_{j}^{\nu}})\Big]
	\,\tfrac{L}{R_{j}}\,, \quad
	\label{div ij} \\
	\partial_{U_{i}}\!\cdot\partial_{U_{j}}\eq
	\partial_{v_{i}}\,Z(x_{i},x_{j})\,\partial_{v_{j}}
	+ \partial_{u_{i}^{\mu}}\,H^{\mu}(x_{i},x_{j})\,\partial_{v_{j}}
	+ \partial_{v_{i}}\,H^{\nu}(x_{j},x_{i})\,\partial_{u_{j}^{\nu}} \nn
	&& +\,\partial_{u_{i}^{\mu}}\,G^{\mu\nu}(x_{i},x_{j})\,\partial_{u_{j}^{\nu}}\,,
	\label{tr ij}
\ea
where we have also introduced the compact notation
\be
	u^{\mu}_{i}:=u_{i}^{\alpha} \ e_{\alpha}^{\ \mu}(x_{i})\,,
	\qquad
	\partial_{u_{i}^{\mu}}:=\partial_{u_{i}^{\alpha}} \ e^{\alpha}_{\ \mu}(x_{i})\,.
\ee
These quantities are more convenient than the flat auxiliary variables for the explicit computations since they commute with the (A)dS-covariant derivative:
\be
	[\,D_{i\,\mu}\,,\, u_{j}^{\nu}\,]=0\,,\qquad	[\,D_{i\,\mu}\,,\, \partial_{u_{j}^{\nu}}\,]=0\,.
\ee
The advantage of the bi-local quantities \eqref{bilocal} rests on the fact that they are closed under the action of the (A)dS-covariant derivatives,
 as one can see by explicit computation:
\ba
	D_{i\,\mu}\,Z(x_{i},x_{j})=\tfrac1L\,H_{\mu}(x_{i},x_{j})\,,&\qquad
	&D_{i\,\nu}\,H_{\mu}(x_{i},x_{j})=-g_{\mu\nu}(x_{i})\,\tfrac1L\,Z(x_{i},x_{j})\,, \nn
	D_{j\,\nu}\,H_{\mu}(x_{i},x_{j})=\tfrac1L\,G_{\mu\nu}(x_{i},x_{j})\,,&\qquad
	&D_{i\,\rho}\,G_{\mu\nu}(x_{i},x_{j})=-g_{\rho\mu}(x_{i})\,\tfrac1L\,H_{\nu}(x_{j},x_{i})\,.
	\qquad
	\label{der bilocal}
\ea
Therefore, the ambient-space cubic vertices \eqref{sol ads} can be reduced to the (A)dS-intrinsic ones with some algebra. Notice that the ambient-space derivatives $\partial_{X_{i}^{\sst M}}$ do not always reduce to
the (A)dS covariant ones $D_{i\,\mu}$\,, but they can produce some powers of $1/L^{2}$ either via the contractions between $\partial_{v_{i}}/L$'s and $v_{i}/L$'s or
via the actions on bi-local quantities.
Hence, an ambient-space vertex with a number $\Delta$
of ambient-space derivatives results in different (A)dS vertices whose number of covariant derivatives
varies within the range \mt{\Delta,\, \Delta-2,\, \cdots ,1} (or $0$).
Whenever the number of derivatives decreases by two,
the corresponding mass-dimension is compensated by a factor $1/L^{2}$\,.

\subsubsection*{Example: 3$-$3$-$2 vertex with lowest number of derivatives}

Let us deal with an explicit example in order to see how this radial reduction works.
We have chosen the 3$-$3$-$2 example with the least number of derivatives because it is both one of the simplest examples of HS interactions and one of the vertices constructed by Fradkin and Vasiliev in the frame-like formalism.
The 3$-$3$-$2 vertex was also obtained in  \cite{Boulanger:2008tg, Zinoviev:2008ck}
in terms of metric-like fields.

For simplicity, we leave aside the Chan-Paton factors and choose $\alpha$ and $\beta$ in a way\footnote{We take the \mt{\e\to0} limit with  \mt{\alpha=1-2\e} and \mt{\beta=1+\e}\,. Even though the second line's last factor in \eqref{sol ads} diverges, it does not matter since we consider the case $n=s_{3}$\,.} that the cubic action has a symmetric form:
\ba
	S^{\sst (3)} \eq -\frac{2}{3}\, g^{\sst 332,2}\,
	\int \frac{d^{d+1}X}L\ \delta\Big(\tfrac{\sqrt{X^{2}}}L-1\Big)\
	 \left(\cG^{\Lambda}\right)^{2}\,
	\partial_{U_{1}}\!\!\cdot\partial_{X_{2}}\,\partial_{U_{2}}\!\!\cdot\partial_{X_{1}}
	\times \nn
	&& \hspace{80pt}\times\,
	\Phi^{\sst (3)}(X_{\sst 1},U_{\sst 1})\
	\Phi^{\sst (3)}(X_{\sst 2},U_{\sst 2})\ \Phi^{\sst (2)}(X_{\sst 3},U_{\sst 3})\,
	\Big|_{\overset{X_{1}=X_{2}=X_{3}=X}{\sst U_{1}=U_{2}=U_{3}=0}}\,,
	\label{ex 332}
\ea
where $\cG^{\Lambda}$ is given by
\be
	\cG^{\Lambda}=2\,\big[
	\partial_{U_{2}}\!\!\cdot\partial_{U_{3}} \partial_{U_{1}}\!\!\cdot\partial_{X_{2}}\,
	-\partial_{U_{1}}\!\!\cdot\partial_{U_{3}}\,\partial_{U_{2}}\!\!\cdot\partial_{X_{1}}
	+ \tfrac12\,\partial_{U_{1}}\!\!\cdot\partial_{U_{2}}\,\partial_{U_{3}}\!\!\cdot\partial_{X_{12}}\big]\,.
	\label{G11}
\ee
Expanding $\left(\cG^{\Lambda}\right)^{2}$ gives rise to six terms,
and in order to describe the procedure let us consider first the term
\be
	(\partial_{U_{2}}\!\!\cdot\partial_{U_{3}})^{2}\,(\partial_{U_{1}}\!\!\cdot\partial_{X_{2}})^{3}\,
	\partial_{U_{2}}\!\!\cdot\partial_{X_{1}}\,.
	\label{1 term}
\ee
Using \eqref{tr ij} and \eqref{div ij}, one gets
\ba
	&&
	\Big[ \partial_{v_{2}}\,Z_{\sst 23}\,\partial_{v_{3}}
	+\partial_{u_{2}}\!\!\cdot H^{\sst 2}_{\sst \ \,3}\,\partial_{v_{3}}\,
	+ \partial_{u_{3}}\!\!\cdot H^{\sst 3}_{\sst \ \,2}\,\partial_{v_{2}}\,
	+ \partial_{u_{2}}\!\!\cdot G^{\sst 23}\!\!\cdot \partial_{u_{3}} \Big]^{2}
	\nn &&
	\times\,\Big[(\partial_{v_{1}}\,Z_{\sst 12}+
	\partial_{u_{1}}\!\!\cdot H^{\sst 1}_{\sst\ \,2})\,
	\partial_{R_{2}} +
	(\partial_{v_{1}}\,H_{\sst 1}^{\sst\ 2}+\partial_{u_{1}}\!\!\cdot G^{\sst 12})\!\cdot\!
	\big[D_{\sst 2}+\tfrac1L\,(u_{\sst 2}\,\partial_{v_{2}}-v_{\sst 2}\,\partial_{u_{2}})\big]
	\,\tfrac{L}{R_{2}}\Big]^{3}
	\nn &&
	\times\,\Big[(\partial_{v_{2}}\,Z_{\sst 21}+
	\partial_{u_{2}}\!\!\cdot H^{\sst 2}_{\sst\ \,1})\,
	\partial_{R_{1}} +
	(\partial_{v_{2}}\,H_{\sst 2}^{\sst\ 1}+\partial_{u_{2}}\!\!\cdot G^{\sst 21})\!\cdot\!
	\big[D_{\sst 1}+\tfrac1L\,(u_{\sst 1}\,\partial_{v_{1}}-v_{\sst 1}\,\partial_{u_{1}})\big]\,
	\tfrac{L}{R_{1}}\Big]\,,\qquad
	\label{intm}
\ea
where the subscripts or superscripts of $Z, H$ and $G$ encode the bi-local dependence on $(x_{i},x_{j})$\,, in particular,
\be
(H^{\sst i}_{\sst\ \,j})_{\mu}=H_{\mu}(x_{i},x_{j})\ ,\qquad (H_{\sst i}^{\sst\ j})_{\mu}=H_{\mu}(x_{j},x_{i})\,.
\ee
Even though eq.~\eqref{intm} has a rather complicated structure, many simplifications can be made. First, since
the operator \eqref{1 term} is acting on
\be
	 \tfrac{R_{1}}L\,\tfrac{R_{2}}L\
	 \varphi^{\sst (3)}(x_{1},u_{1})\,\varphi^{\sst (3)}(x_{2},u_{2})\,
	 \varphi^{\sst (2)}(x_{3},u_{3})\,\big|_{u_{i}=v_{i}=0}\,,
	 \label{hom cond}
\ee
the dependence in $R_{i}$ and $v_{i}$  can be removed
performing all possible contractions.
Second, the coincident limit \eqref{prop bilocal} simplifies some of the bi-local quantities, and the formula \eqref{intm} becomes
\ba
	&& (\partial_{u_{2}}\!\!\cdot\partial_{u_{3}})^{2}
	\,\Big[\, \Big\{ (\partial_{u_{1}}\!\!\cdot D_{\sst 2})^{2}
	(\partial_{u_{1}}\!\!\cdot G^{\sst 12}\!\!\cdot D_{\sst 2}+\tfrac1L\,\partial_{u_{1}}\!\!\cdot H^{\sst 1}_{\sst\ \,2})
	-\tfrac3{L^{2}}\,u_{2}\!\cdot\partial_{u_{1}}\,\partial_{u_{1}}\!\!\cdot\partial_{u_{2}}\,
	\partial_{u_{1}}\!\!\cdot D_{\sst 2}\Big\}\times\nn
	&&\hspace{65pt} \times\,
	(\partial_{u_{2}}\!\!\cdot G^{\sst 21}\!\!\cdot D_{\sst 1}
	+\tfrac1L\,\partial_{u_{2}}\!\!\cdot H^{\sst 2}_{\sst\ \,1}) \nn
	&&\hspace{58pt} +\,\tfrac1{L}\,\Big\{
	\tfrac1L\,(\partial_{u_{1}}\!\!\cdot D_{\sst 2})^{2}
	-\partial_{u_{1}}\!\!\cdot D_{\sst 2}\ H_{\sst 1}^{\sst\ 2}\!\cdot D_{\sst 2}\,
	(\partial_{u_{1}}\!\!\cdot G^{\sst 12}\!\!\cdot D_{\sst 2}+\tfrac1L\,\partial_{u_{1}}\!\!\cdot H^{\sst 1}_{\sst\ \,2})
	 \nn
	&&\hspace{86pt} +\,\tfrac1{L^{2}}\,\partial_{u_{1}}\!\!\cdot D_{\sst 2}\ u_{2}\!\cdot H^{\sst 2}_{\sst\ \,1}\,
	\partial_{u_{1}}\!\!\cdot G^{\sst 12}\!\!\cdot \partial_{u_{2}}
	-(\partial_{u_{1}}\!\!\cdot D_{\sst 2})^{2}(H_{\sst 1}^{\sst\ 2}\!\cdot D_{\sst 2}+\tfrac1L\,Z_{\sst 12}) \nn
	&& \hspace{86pt}+\,
	\tfrac2{L^{2}}\,u_{2}\!\cdot\partial_{u_{1}}\,\partial_{u_{1}}\!\!\cdot D_{\sst 2}\
	\partial_{u_{2}}\!\!\cdot H^{\sst 2}_{\sst\ \,1}\, \Big\}\,
	\partial_{u_{1}}\!\!\cdot G^{\sst 12}\!\!\cdot \partial_{u_{2}}\,\Big]\,.
	\label{332 1}
\ea
Finally, the property \eqref{der bilocal} enables one to remove all bi-local quantities
replacing them with some powers of $L$\,. At the end, one obtains the (A)dS intrinsic expression for the operator \eqref{1 term} as
\be
	(\partial_{u_{2}}\!\!\cdot\partial_{u_{3}})^{2}\,
	(\partial_{u_{1}}\!\!\cdot D_{\sst 2})^{3}\,
	\partial_{u_{2}}\!\!\cdot D_{\sst1}
	-\tfrac6{L^{2}}\,
	\partial_{u_{1}}\!\!\cdot\partial_{u_{2}}\,
	\partial_{u_{1}}\!\!\cdot\partial_{u_{3}}\,
	\partial_{u_{2}}\!\!\cdot\partial_{u_{3}}\,
	\partial_{u_{1}}\!\!\cdot D_{\sst 2}\,\partial_{u_{2}}\!\!\cdot D_{\sst1}\,.
	\label{332 1 ds}
\ee
Notice that the first term has the same form of \eqref{1 term} with the replacement of $(\partial_{X_{i}}\,,\partial_{U_{i}})$ by $(D_{\sst i}\,,\,\partial_{u_{i}})$, but the second term has a lower number of derivatives and is proportional to $1/L^{2}$\,.

Five other terms in the expansion \eqref{ex 332} can be computed in a similar way (see Appendix \ref{sec:reduc} for more details),
and the cubic action \eqref{ex 332} can be finally expressed solely in terms of (A)dS intrinsic quantities as
\ba
	&& S^{\sst (3)} = -\frac{8}{3}\, g^{\sst 332,2}\,
	\int d^{d}x\,\sqrt{-g}\,\times \nn
	&& \times\,\Big[
	\left(
	\partial_{u_{2}}\!\!\cdot\partial_{u_{3}}\,\partial_{u_{1}}\!\!\cdot D_{\sst 2}
	-\partial_{u_{1}}\!\!\cdot\partial_{u_{3}}\,\partial_{u_{2}}\!\!\cdot D_{\sst 1}
	+\tfrac12\,\partial_{u_{1}}\!\!\cdot\partial_{u_{2}}\,\partial_{u_{3}}
	\!\!\cdot D_{\sst 12}\right)^{2}
	\partial_{u_{1}}\!\!\cdot D_{\sst 2}\,
	\partial_{u_{2}}\!\!\cdot D_{\sst 1}\nn
	&& \quad\
	+\,\tfrac4{L^{2}}\,\partial_{u_{1}}\!\!\cdot\partial_{u_{2}}
	\left[
	(\partial_{u_{2}}\!\!\cdot\partial_{u_{3}})^{2}
	(\partial_{u_{1}}\!\!\cdot D_{\sst 2})^{2}
	+
	(\partial_{u_{1}}\!\!\cdot\partial_{u_{3}})^{2}
	(\partial_{u_{2}}\!\!\cdot D_{\sst 1})^{2}
	-3\,\partial_{u_{1}}\!\!\cdot\partial_{u_{3}}\,\partial_{u_{2}}\!\!\cdot\partial_{u_{3}}
	\,\partial_{u_{1}}\!\!\cdot D_{\sst 2}\,\partial_{u_{2}}\!\!\cdot D_{\sst 1}
	\right] \nn
	&& \quad\
	+\,\tfrac3{L^{2}}\,(\partial_{u_{1}}\!\!\cdot\partial_{u_{2}})^{2}\,
	\left[	\partial_{u_{2}}\!\!\cdot\partial_{u_{3}}\,
	\partial_{u_{1}}\!\!\cdot D_{\sst 2}
	-\partial_{u_{1}}\!\!\cdot\partial_{u_{3}}\,
	\partial_{u_{2}}\!\!\cdot D_{\sst 1}
	+\tfrac16\,\partial_{u_{1}}\!\!\cdot\partial_{u_{2}}\,
	\partial_{u_{3}}\!\!\cdot D_{\sst 12}
	\right] \partial_{u_{3}}\!\!\cdot D_{\sst 12}
	\nn
	&& \quad\
	-\,\tfrac5{4\,L^{2}}\,(\partial_{u_{1}}\!\!\cdot\partial_{u_{2}})^{2}\,
	\left[	\partial_{u_{2}}\!\!\cdot\partial_{u_{3}}\,
	\partial_{u_{1}}\!\!\cdot D_{\sst 2}\,\partial_{u_{3}}\!\!\cdot D_{\sst 1}
	+\partial_{u_{1}}\!\!\cdot\partial_{u_{3}}\,
	\partial_{u_{2}}\!\!\cdot D_{\sst 1}\,\partial_{u_{3}}\!\!\cdot D_{\sst 2}
	\right] 	\nn
	&&\quad\
	-\,\tfrac{7d+29}{2\,L^{4}}\,
	(\partial_{u_{1}}\!\!\cdot\partial_{u_{2}})^{2}
	\partial_{u_{1}}\!\!\cdot\partial_{u_{3}}\,\partial_{u_{2}}\!\!\cdot\partial_{u_{3}}
	\Big]\ \varphi^{\sst (3)}(x_{\sst1},u_{\sst 1})\,
	\varphi^{\sst (3)}(x_{\sst 2},u_{\sst 2})\,\varphi^{\sst (2)}(x_{\sst 3},u_{\sst 3})\,
	\Big|_{\overset{x_{1}=x_{2}=x_{3}=x}{\sst u_{1}=u_{2}=u_{3}=0}}\,,
	\label{ex 332 dS}
\ea
where we organized the various contributions according to
the number of (A)dS covariant derivatives.
%


\section{Discussion and Outlook}
\label{sec: discussions}

In this paper, we have obtained the general solution of the cubic-interaction problem of HS gauge fields in (A)dS within the setting of the TT constraints
that we have defined in the Introduction.
Interestingly, the structure of the vertices, when expressed in the ambient-space formalism, coincides with the flat-space ones up to non-trivial total-derivative terms whose form is completely constrained by the gauge consistency.

\subsection{Relation to the Fradkin-Vasiliev vertices}

Let us consider the \mt{s\!-\!s\!-\!2} vertices, which
were originally constructed by Fradkin and Vasiliev (FV).
They correspond to the case \mt{s_{1}=s_{2}=s} and \mt{s_{3}=n=2} in \eqref{sol ads}, so that they are the vertices with lowest number of derivatives.
With the same choice of $\alpha$ and $\beta$ as in \eqref{ex 332},
they are given by
\ba
	S^{\sst (3)} \eq g^{ss{\sst 2,2}}\,
	\int \frac{d^{d+1}X}L\ \delta\Big(\tfrac{\sqrt{X^{2}}}L-1\Big)\
	 \left(\cG^{\Lambda} \right)^{2}\,
	(\partial_{U_{1}}\!\!\cdot\partial_{X_{2}}\,\partial_{U_{2}}\!\!\cdot\partial_{X_{1}})^{s-2}
	\times \nn
	&& \hspace{100pt}\times\,
	\Phi^{\sst (s)}(X_{\sst 1},U_{\sst 1})\
	\Phi^{\sst (s)}(X_{\sst 2},U_{\sst 2})\ \Phi^{\sst (2)}(X_{\sst 3},U_{\sst 3})\,
	\Big|_{\overset{X_{1}=X_{2}=X_{3}=X}{\sst U_{1}=U_{2}=U_{3}=0}}\,,\qquad
	\label{ss2}
\ea
where, for simplicity, we have absorbed a numerical factor into the definition of
the coupling constant. In the previous section, we have shown how to express ambient differential operators in terms of (A)dS-intrinsic quantities.
Likewise, expressing the operators
\mt{(\partial_{U_{1}}\!\!\cdot\partial_{X_{2}}\,\partial_{U_{2}}\!\!\cdot\partial_{X_{1}})^{s-2}}
 in the above formula using \eqref{div ij} yields an expression in terms of (A)dS-covariant derivatives, bi-local quantities and also the $v_{i}$'s. Taking the ordering where all
 (A)dS-covariant derivatives are placed on the RHS, one gets
\be
	(\partial_{U_{1}}\!\!\cdot\partial_{X_{2}}\,\partial_{U_{2}}\!\!\cdot\partial_{X_{1}})^{s-2}
	=\cA_{s-2}+\Lambda\,\cA_{s-3}+\cdots+\Lambda^{s-2}\,\cA_{0}\,,
	\label{A exp}
\ee
where $\cA_{r}$ is the portion containing the $2r$-th power of the (A)dS-covariant derivatives, or $\cA_{r}\propto D^{2r}$\,. Plugging \eqref{A exp} into \eqref{ss2}, the \mt{s\!-\!s\!-\!2} vertex admits a similar expansion given by
\be
	S^{\sst (3)} = g^{ss{\sst 2,2}}\left[
	A_{s}+\Lambda\,A_{s-1}+\cdots+\Lambda^{s-2}\,A_{2} \right],
	\label{S exp}
\ee
with
\be
	A_{r+2} = \int \frac{d^{d+1}X}L\ \delta\Big(\tfrac{\sqrt{X^{2}}}L-1\Big)\,
	 \left(\cG^{\Lambda}\right)^{2}\,\cA_{r}\
	\Phi^{\sst (s)}\, \Phi^{\sst (s)}\, \Phi^{\sst (2)}\,
	\Big|_{\overset{X_{1}=X_{2}=X_{3}=X}{\sst U_{1}=U_{2}=U_{3}=0}}\,.
\ee
Notice that each $A_{r}$ is separately gauge invariant under the spin 2 gauge transformation,
and this is due to the fact that the $\cA_{r}$'s trivially commute with the spin 2 gauge transformation.
Notice as well that $A_{r}$ involves $2(r-2)$ or $2(r-1)$ (A)dS covariant derivatives,
since the action of $ \left(\cG^{\Lambda}\right)^{2}$ may or may not add two additional derivatives.

This expansion of the vertex is quite similar to the one obtained by FV, and in fact
one can make it as an expansion in inverse powers of $\Lambda$
by redefining the coupling constant and the fields as
\be
	g^{ss{\sst 2,2}}=\frac{\sqrt{G}}{\Lambda^{s-2}}\ \lambda_{s}\,, \qquad
	\varphi^{\sst (2)}=\tfrac1{\sqrt{G}}\,h\,,
	\qquad
	\varphi^{\sst (s)}=\tfrac1{\sqrt{G}}\,\phi^{\sst (s)}\,.
	\label{redef F}
\ee
The coupling constant $g^{ss{\sst 2,2}}$ has mass-dimension $(2-d)/2-2(s-2)$\,,
while with this redefinition the new coupling constant $\lambda_{s}$ together with the new fields
have vanishing mass-dimension.
Finally, the expansion \eqref{S exp} becomes
\be
	S^{\sst (3)} = \frac{\lambda_{s}}G\left[ \tilde A_{2}
	+\frac1\Lambda\, \tilde A_{3}+\cdots+\frac1{\Lambda^{s-2}}\, \tilde A_{s} \right],
\ee
where $\tilde A_{r}$'s are given schematically by eq.~\eqref{FV ver}
in terms of dimensionless fields $\phi^{\sst (s)}$ and $h$\,.

Two remarks are in order. First, the lowest-derivative part $\tilde A_{2}$ of the above expression
should involve the gravitational minimal coupling as well as non-minimal ones which do not
deform the gauge transformation.
Therefore, the simplest way to see this link is to analyze how the vertices constructed in this paper
deform the gauge transformation and the gauge algebra.
We leave this issue for future work.
Second, the highest-derivative part
(the so-called \emph{seed coupling}, according to \cite{Sagnotti:talk09})
has the same form as
the flat-space vertices with $\partial_{x_{i}^{\mu}}$'s replaced by $D_{i\,\mu}$'s.
The relation between the gravitational minimal coupling and the seed coupling was already noticed in \cite{Boulanger:2008tg},
and in the present work we can see how
both lower-derivative and seed couplings come out at the same time from the ambient-space vertices.
From a more general perspective, it would be interesting to investigate the relation between the present construction (in metric-like approach) and the recent frame-like one \cite{Vasiliev:2011xf}.

\paragraph{Boulanger-Leclercq-Sundell limit}

Since a curved space looks flat in the short-distance limit, the dominant term of curved-space actions in the limit should correspond to flat-space ones. One may expect to obtain in this way the flat-space vertices from the FV ones, but because of the inverse power expansion in $\Lambda$
the dominant terms diverge in the limit. In \cite{Boulanger:2008tg}, the authors considered a particular limit of the FV system in order to extract flat-space information from AdS interactions. More precisely, they considered the limit where not only the cosmological constant but also the gravitational constant and fields scale as
\ba
	&\Lambda=\epsilon\,\tilde\Lambda\,,\qquad
	&G=\epsilon^{2(s-2)}\,\tilde G\,,\nn
	& h=\epsilon^{s-2}\,\tilde h\,,\qquad
	& \phi^{\sst (s)}= \epsilon^{s-2}\,\tilde\phi^{\sst (s)}\,,
	\label{BLS}
\ea
with $\epsilon\to0$\,. Under this rescaling, the quadratic action remains invariant, but the cubic vertices scale in a way that only the seed coupling survives and one gets the flat-space vertices with $2s-2$ derivatives.

In our setting, this can be understood at the level of \eqref{S exp}, where the flat-space limit is not singular for fixed (or non-scaling) $g^{ss{\sst 2,2}}$ and $\varphi^{\sst (s)}$'s, and the flat-space cubic vertices are recovered. In this respect, the rescaling \eqref{BLS} can be viewed as a particular flat-space limit in \eqref{redef F} which holds $g^{ss{\sst 2,2}}$ and $\varphi^{\sst (s)}$'s finite.

\subsection{Outlook}

Let us conclude by summarizing
our results and strategies of the present paper and
by describing their possible applications
from a more general perspective.

First, we observed that
the flat-space interactions play a key role,
through the ambient-space formalism,
in understanding and controlling cubic interactions in any constant curvature background.
Second,
the simplified TT (or \emph{on-shell}) system makes possible to
identify the consistent cubic interactions
and this observation has recently led to the proposal \cite{Taronna:2011kt}
for quartic and higher-order HS interactions by one of the authors.
We expect as well that many other key properties of the interactions can be appreciated at this simpler level.

The aforementioned perspectives open a new window for a systematic analysis of many other aspects of HS theory.
First of all, the issue of non-Abelian HS gauge algebras in (A)dS and flat space together with their relations might be addressed\footnote{
See  \cite{Bekaert:2010hp} for an analysis of flat-space gauge algebras.}.
In particular, it is interesting to draw some more lessons on the HS geometry\footnote{
See \cite{Francia:2002aa,Francia:2002pt} for the free HS geometric equations, and \cite{Manvelyan:2010jf} for a recent development on HS curvatures.}
from the relations between the \emph{minimal} (A)dS couplings
and \emph{non-minimal} ambient-space ones.
Moreover, the nature of non-localities, which appear in the flat-space Lagrangian starting from the quartic order,
can be clarified  from this point of view:
flat-space non-localities might fit within the Vasiliev's system with the aid of the ambient-space formalism.
If so,
the strategy employed in this work can
give an additional motivation for the flat-space HS gauge theory.

Further interesting applications of our results can be found in massive HS field theories,\footnote{
See \cite{Francia:2007ee, Francia:2008ac} for the recent development of the free massive HS theory in the metric-like approach.}
of which String Theory is the most important example.
Actually, the interactions of massive HS fields can be investigated with techniques
similar to those used in this paper, and we hope to address on this question soon.
It is indeed believed by many authors\footnote{
See e.g. \cite{Zinoviev:2008jz, Zinoviev:2008ck, Zinoviev:2009hu} for some investigation along these lines.}
that the masses of HS fields can play a role similar to that of the cosmological constant of massless HS theories,
and the understanding of this relation can give more insights
on the very nature of
String Theory.\footnote{
See \cite{Bengtsson:1986ys, Heanault:book, Sagnotti:2003qa}
for the triplet system which contains the same DoFs as
the massless limit of the first Regge trajectory of String Theory.
See \cite{Porrati:2010hm} for the analysis of HS interactions in a constant electromagnetic background within the String Theory framework.
See \cite{Polyakov:2009pk,Polyakov:2010qs,Polyakov:2010sk}
for the construction of some cubic and quartic flat-space vertices of massless HS fields
using vertex operators in String theory, and
\cite{Polyakov:2011sm} for its recent extension to AdS.
}

Moreover, other applications can be found in the AdS/CFT correspondence,
which
has been applied to HS theories starting from \cite{Sezgin:2002rt, Klebanov:2002ja}.\footnote{
In the $AdS_{3}$ case,
there has been considerable recent development after the works \cite{Campoleoni:2010zq, Henneaux:2010xg}.}
We expect that the ambient-space representation of interacting vertices
simplifies the computations of $n$-point functions.
Moreover, loop computations might be addressed within this formalism,
shedding some light on the quantum aspects of HS gauge theories.
We hope to report on these issues in the near future.

\acknowledgments{
We are grateful to
X. Bekaert,
N. Boulanger,
A. Campoleoni,
L. Lopez,
J. Mourad,
P. Sundell
and
M. Tsulaia for helpful conversations,
and especially to
A. Sagnotti for the reading of the manuscript and for many suggestions.
The present research was supported in part
by Scuola Normale Superiore,
by INFN and
by the MIUR-PRIN contract 2009-KHZKRX.
}

\appendix

\section{Foregoing the traceless and transverse constraints}
\label{sec: forego}

In this appendix, we briefly discuss how the TT constraints can be
forgone in a systematic fashion \cite{Manvelyan:2010jr,Sagnotti:2010at}.
We follow the discussion of the flat-space case,
presented in Appendix B of \cite{Taronna:2011kt},
and sketch the generalization of the logic to (A)dS.

Let us first recall the flat-space case.
The idea is to start from the on-shell action and to
compensate the outcome of the gauge variation
that vanished with the TT constraints.
For the quadratic part,
one can see in this way that the action \eqref{flat q act}
can be completed as
\be
S^{\sst (2)}= \frac12\,\int d^{d}x\ e^{\partial_{u_1}\!\cdot\,\partial_{u_2}}
\left(\partial_{x_{2}}^{2}+\cD_{\sst 1}\,\cD_{\sst 2}-
\partial_{x_{2}}^{2}\,\cT_{\sst 1}\,\cT_{\sst 2}\right)
\varphi(x_1,u_1)\,
\varphi(x_2,u_2)\,\big|_{\overset{x_{1}=x_{2}=x}{\sst u_{1}=u_{2}=0}}\,,
\label{act app}
\ee
where the operators $\cD$ and $\cT$ are defined as solutions of the following equations:
\be
\cD\ \d^{\sst (0)}_{\sst \varepsilon}\,\varphi(x,u)=\partial_x^{2}\,\varepsilon(x,u)\,,
\qquad
\cT\ \d^{\sst (0)}_{\sst \varepsilon}\,\varphi(x,u)=
-\,\partial_x\!\cdot\partial_u\,\varepsilon(x,u)\,,
\label{quadsoluz}
\ee
while $\cD$ can be solved in terms of $\cT$ as
\be
\cD = \partial_x\!\cdot\partial_u + u\cdot\partial_x\,\cT\,.
\ee
The latter can be now obtained solving for the operator $\cT$ that has to be found restricting the attention to operators that vanish with the TT constraints.
One can recover the usual Fronsdal formulation \cite{Fronsdal:1978rb}
if we search for a solution for $\cT$ built from traces without allowing for auxiliary fields.
Indeed, at the end one recovers the following solution for $\cT$\,:
\be
\cT=-\tfrac{1}{2}\, \partial_u^{2}\,,
\ee
that satisfies eq.~\eqref{quadsoluz} up to terms that are proportional to the trace of the gauge parameter.
Since the gauge parameter is traceless,
the double trace part of the field is gauge invariant.
Hence, in order to be sure that after gauge fixing the original TT system is recovered,
the double trace of the field must be set to zero.
Let us mention that the different solutions for $\cT$ provide
other possible off-shell completions with the possibility of having different spectra.
For instance, both the compensator
system \cite{Francia:2005bu} and  the triplet system
\cite{Bengtsson:1986ys, Heanault:book, Sagnotti:2003qa,
Francia:2010ap, Francia:2010qp}
can be recovered.

Let us now recall briefly how this procedure works for the flat-space cubic action.
In order to simplify the discussion, we consider the simpler case in which the on-shell cubic coupling function is an exponential
without any  Chan-Paton factor:
\be
\cK=e^{\ell\,\cV}\,, \qquad
\cV=\cY_{1}+\cY_{2}+\cY_{3}+\cG\,,
\ee
where $\cY_{i}$'s and $\cG$ are defined in \eqref{Y&Z}, and $\ell$ is a length parameter.
The basic viewpoint is similar to that considered at the quadratic level and one needs to find the right counterterms to cure the non-zero gauge variation of the cubic
action that is proportional to divergences.
Moreover, gauge invariance should hold up to the Fronsdal equation:
\be
	(\partial_x^{2}-u\cdot\partial_x\,\cD)\,\varphi \approx 0\,,
\ee
since terms proportional to it can be compensated,
and one obtains the following commutation relation
and its cyclic permutations:
\ba
&&
\big[\,\cV\,,\,u_{\sst 1}\!\cdot\partial_{x_1}\big]\,
\varphi_{\sst 1}\,\varphi_{\sst 2}\,\varepsilon_{\sst 3}
\approx
\left(\cH_{23}\ u_{\sst 3}\!\cdot\partial_{x_3}-
u_{\sst 2}\!\cdot\partial_{x_2}\,\cH_{32}\right)
\varphi_{\sst 1}\,\varphi_{\sst 2}\,\varepsilon_{\sst 3}\,,\nn
&&
\big[\,\cV\,,\,u_{\sst 2}\!\cdot\partial_{x_2}\big]\,
\varphi_{\sst 1}\,\varphi_{\sst 2}\,\varepsilon_{\sst 3}
\approx
\left(\cH_{13}\ u_{\sst 1}\!\cdot\partial_{x_1}-
u_{\sst 3}\!\cdot\partial_{x_3}\,\cH_{31}\right)
\varphi_{\sst 1}\,\varphi_{\sst 2}\,\varepsilon_{\sst 3}\,,\nn
&&
\big[\,\cV\,,\,u_{\sst 3}\!\cdot\partial_{x_3}\big]\,
\varphi_{\sst 1}\,\varphi_{\sst 2}\,\varepsilon_{\sst 3}
\approx
\left(u_{\sst 2}\!\cdot\partial_{x_2}\ \cH_{12}-
u_{\sst 1}\!\cdot\partial_{x_1}\,\cH_{21}\right)
\varphi_{\sst 1}\,\varphi_{\sst 2}\,\varepsilon_{\sst 3}\,,\nn
\ea
where
\be
	\cH_{ij}=(\partial_{u_{i}}\!\!\cdot\partial_{u_{j}}+1) \,\cD_{j}
	+\partial_{u_{i}}\!\!\cdot\partial_{x_{j}}\,\cT_{j}\,.
\ee
This set of commutators
encodes a recursive structure from which one can reconstruct the full \emph{off-shell} completion of the cubic vertex. The end result can be  expressed in the following compact form:
\be
{\cK}^{\sst \,\text{off-shell}}=
e^{\ell\, \cV}\,
\Big[1+\ell^{2}\ {\cH}_{12}\,{\cH}_{13}
+\ell^{3}\,:{\cH}_{21}\,{\cH}_{32}\,{\cH}_{13}:\,
+\ \text{(cyclic\ perm.)} \Big]\,,
\label{genoffshell}
\ee
where $:\ :$ enforces an ordering in which the generalized de Donder operators
are to act directly on the fields and hence are to be put to the right:
\be
:\,{\cD}_{1}\ \cZ_{2}:\ =\cZ_{2}\ {\cD}_{1}\,.
\ee

The (A)dS off-shell completion is expected to work on very similar grounds. In the following, for simplicity, we consider the quadratic part and then sketch how this procedure extends to
the cubic level. The only complication of (A)dS is the presence of the $\d$-function in the action, that requires in principle further counterterms in order to compensate
for the various total derivatives that are generated.
The starting point is now eq.~\eqref{freeAdS} and the only additional terms
in the gauge variation
that are to be considered are those proportional to a total derivative and
are of the form:
\be
\d\, S^{\sst (2)}=\int \frac{d^{d+1}X}L\ \delta\Big(\tfrac{\sqrt{X^{2}}}L-1\Big)\
\partial_{X}\!\cdot\partial_{U_{\sst 2}}\,
e^{\partial_{U_1}\!\cdot\,\partial_{U_2}}\,
\partial_{X_2}^{2}\,
E(X_{\sst1}, U_{\sst1})\,\Phi (X_{\sst 2}, U_{\sst 2})\,
\Big|_{\overset{X_{1}=X_{2}=X}{\sst U_{1}=U_{2}=0}}\,.
\ee
This additional term, after an integration by parts,
gives $X_{1}\!\cdot\partial_{U_{1}}$ and vanishes with
the tangent condition \eqref{ht con}.
Hence, the off-shell ambient-space action for totally symmetric (A)dS fields is
\begin{multline}
S^{\sst (2)} = \frac1{2} \int \frac{d^{d+1}X}L\ \delta\Big(\tfrac{\sqrt{X^{2}}}L-1\Big)\,\times
\\
\times\,e^{\partial_{U_1}\!\cdot\,\partial_{U_2}}
\left(\partial_{X_{2}}^{2}+\cD_{\sst 1}\,\cD_{\sst 2}-
\partial_{X_{2}}^{2}\,\cT_{\sst 1}\,\cT_{\sst 2}\right)
\Phi(X_1,U_1)\,
\Phi(X_2,U_2)\,\Big|_{\overset{X_{1}=X_{2}=X}{\sst U_{1}=U_{2}=0}}\,.
\end{multline}
The same procedure that we have considered in the simple quadratic case has to be performed also at the cubic order.
Starting from the ambient-space analogue of
the off-shell cubic vertex $S^{\sst (3)
}$ \eqref{genoffshell}, one gets
the following schematic form of the gauge variation
\begin{multline}
	\delta\,S^{\sst (3)}=
	\int d^{d+1}X\ \sum_{n=1}^{\infty}\,
	\delta^{\sst (n)}\Big(\tfrac{\sqrt{X^{2}}}L-1\Big)\ \times\\
	\times\,
	\partial_{X_{i}}\!\!\cdot\partial_{U_{i}}\,\big( \cdots \big)\,
	E(X_{\sst1}, U_{\sst1})\,\Phi (X_{\sst 2}, U_{\sst 2})\,\Phi (X_{\sst 3}, U_{\sst 3})
	\Big|_{\overset{X_{1}=X_{2}=x_{3}=X}{\sst U_{1}=U_{2}=U_{3}=0}}\,,
\end{multline}
that is to be compensated adding further divergence and trace terms
at the $\delta^{\sst (1)}$-level.
This procedure is expected to work order by order, so leading eventually to
the off-shell form of the (A)dS cubic action.

\section{Proof at the $\delta^{\sst (2)}$ level}
\label{sec:C=0}

In this section, we prove that the total-derivative part $\cC$ in \eqref{gv end} does not impose additional conditions on the constants $\alpha_{i}$ and $\beta_{i}$\,. At the level of $\delta^{\sst (1)}$, $\cC$ does not vanish  with (\ref{B1} - \ref{B2}), but is simplified to
\ba
	&& \cC= ({\alpha_{1}}^{2}-1)\,\partial_{X}\!\cdot\partial_{U_{1}}
	+2(\alpha_{2}+1)\,\partial_{X}\!\cdot\partial_{U_{2}}
	-2(\alpha_{3}-1)\,\partial_{X}\!\cdot\partial_{U_{3}}
	\nn
	&&\qquad
	-\,(\alpha_{1}-1)(\beta_{3}-\tfrac12)\,\partial_{X}\!\cdot\partial_{U_{2}}\,\partial_{U_{3}}\!\!\cdot\partial_{U_{1}}
	-(\alpha_{1}+1)(\beta_{2}+\tfrac12)\,\partial_{X}\!\cdot\partial_{U_{3}}\,\partial_{U_{1}}\!\!\cdot\partial_{U_{2}}
	\nn
	&&\qquad
	+\,\big[\,
	2(\alpha_{1}\,\beta_{1}-1)\,\partial_{X}\!\cdot\partial_{U_{1}}
	+\tfrac32(\alpha_{2}+1)\,\partial_{X}\!\cdot\partial_{U_{2}}
	-\tfrac32(\alpha_{3}-1)\,\partial_{X}\!\cdot\partial_{U_{3}}\,\big]\,\partial_{U_{2}}\!\!\cdot\partial_{U_{3}}
	\nn
	&&\qquad
	+\,\big[\,
	({\beta_{1}}^{2}-1)\,\partial_{X}\!\cdot\partial_{U_{1}}\,\partial_{U_{2}}\!\!\cdot\partial_{U_{3}}
	-(\beta_{3}-\tfrac12)(\beta_{1}+\beta_{2})
	\,\partial_{X}\!\cdot\partial_{U_{2}}\,\partial_{U_{3}}\!\!\cdot\partial_{U_{1}} \nn
	&&\qquad\qquad
	-\,(\beta_{2}+\tfrac12)(\beta_{3}+\beta_{1})
	\,\partial_{X}\!\cdot\partial_{U_{3}}\,\partial_{U_{1}}\!\!\cdot\partial_{U_{2}}\,\big]\,
	\partial_{U_{2}}\!\!\cdot\partial_{U_{3}}\,.
\ea
We integrate by parts in order to replace \mt{\delta^{\sst (1)}\,\partial_{X}\!\cdot\partial_{U_{i}}}
with \mt{-\delta^{\sst(2)}\,X_{i}\!\cdot\partial_{U_{i}}/L^{2}}\,,
and then $\delta^{\sst (1)}\,\cC$ can be rewritten as $-\delta^{\sst (2)}\,\cD/L^{2}$
with $\cD$ some other differential operator.
We now push $\cD$ to the right hand side of $e^{L\,\cV}$ as
\be
	\int\ \delta^{\sst (2)}\ k\ \cD\ e^{L\,\cV}\ E_{1}\,\Phi_{2}\,\Phi_{3}\,\big|
	= \int\ \delta^{\sst (2)}\ k\ e^{L\,\cV}\ L\ \cE\ E_{1}\,\Phi_{2}\,\Phi_{3}\,\big|\,,
\ee
getting the following operator acting on the fields:
\ba
	&& \cE= \Big\{-\big[\,2(\alpha_{1}\,\beta_{1}-1)(\alpha_{2}-1)
	-(\alpha_{1}+1)(\alpha_{2}+1)(\beta_{2}-1)+2(\alpha_{2}+1)(\beta_{3}+\beta_{1})\,\big]\,
	\partial_{U_{1}}\!\!\cdot\partial_{U_{2}} \nn
	&&\hspace{32pt}
	-\,\big[\,2(\alpha_{1}\,\beta_{1}-1)(\alpha_{3}+1)-(\alpha_{3}-1)(\alpha_{1}-1)(\beta_{3}+1)
	+2(\alpha_{3}-1)(\beta_{1}+\beta_{2})\,\big]\,\partial_{U_{3}}\!\!\cdot\partial_{U_{1}}\nn
	&&\hspace{32pt}
	-\,\big[\,-(\alpha_{2}-1)({\beta_{1}}^{2}-1)+(\alpha_{2}+1)(\beta_{2}-1)(\beta_{3}+\beta_{1})\,\big]\,
	\partial_{U_{1}}\!\!\cdot\partial_{U_{2}}\,\partial_{U_{2}}\!\!\cdot\partial_{U_{3}} \nn
	&& \hspace{32pt}
	-\,\big[\,-(\alpha_{3}+1)({\beta_{1}}^{2}-1)+(\alpha_{3}-1)(\beta_{3}+1)(\beta_{1}+\beta_{2})\,\big]\,
	\partial_{U_{3}}\!\!\cdot\partial_{U_{1}}\,\partial_{U_{2}}\!\!\cdot\partial_{U_{3}} \nn
	&&\hspace{32pt}
	+\,(\beta_{2}+\beta_{3})\,\big[\,
	\alpha_{1}(\beta_{2}+\beta_{3})+\beta_{2}-\beta_{3}+2\,\big]\,
	\partial_{U_{1}}\!\!\cdot\partial_{U_{2}}\,\partial_{U_{3}}\!\!\cdot\partial_{U_{1}} \Big\}
	\,\partial_{U_{2}}\!\!\cdot\partial_{U_{3}}\,.
\ea
None of these contributions can be compensated,
so that each coefficient in the above formula should vanish separately.
Using the general solutions \eqref{sol cnst} of (\ref{B1} - \ref{B3}), one can
verify that this is indeed the case.

\section{Radial reduction of the 3$-$3$-$2 vertex}
\label{sec:reduc}

In this Appendix we present more details of the reduction of the 3$-$3$-$2 vertex \eqref{ex 332} to the (A)dS-intrinsic expression \eqref{ex 332 dS}.
Expanding the operator in eq.~\eqref{ex 332}
gives altogether six terms:
\ba
	&&
	\big[ \partial_{U_{2}}\!\!\cdot\partial_{U_{3}}\,\partial_{U_{1}}\!\!\cdot\partial_{X_{2}}
	-\partial_{U_{1}}\!\!\cdot\partial_{U_{3}}\,\partial_{U_{2}}\!\!\cdot\partial_{X_{1}}\,
	+ \tfrac12\,\partial_{U_{1}}\!\!\cdot\partial_{U_{2}}
	\,\partial_{U_{3}}\!\!\cdot\partial_{X_{12}}\big]^{2}
	 \,\partial_{U_{1}}\!\!\cdot\partial_{X_{2}}\,\partial_{U_{2}}\!\!\cdot\partial_{X_{1}} \nn
	&& =\,\partial_{U_{2}}\!\!\cdot\partial_{U_{3}}\,
	(\partial_{U_{1}}\!\!\cdot\partial_{X_{2}})^{3}\,\partial_{U_{2}}\!\!\cdot\partial_{X_{1}}\,
	 + (1\leftrightarrow2)\nn
	&&\quad-\,2\,
	\partial_{U_{1}}\!\!\cdot\partial_{U_{3}}\,\partial_{U_{2}}\!\!\cdot\partial_{U_{3}}\,
	(\partial_{U_{1}}\!\!\cdot\partial_{X_{2}})^{2}\,(\partial_{U_{2}}\!\!\cdot\partial_{X_{1}})^{2}
	\nn
	&& \quad+\,
	\partial_{U_{1}}\!\!\cdot\partial_{U_{2}}\,
	\partial_{U_{2}}\!\!\cdot\partial_{U_{3}}\,
	(\partial_{U_{1}}\!\!\cdot\partial_{X_{2}})^{2}\,\partial_{X_{1}}\!\!\cdot\partial_{U_{2}}\,
	\partial_{U_{3}}\!\!\cdot\partial_{X_{12}} + (1\leftrightarrow2)\nn
	&& \quad+\,\tfrac14\,
	 (\partial_{U_{1}}\!\!\cdot\partial_{U_{2}})^{2}\,
	\partial_{U_{1}}\!\!\cdot\partial_{X_{2}}\,
	\partial_{U_{2}}\!\!\cdot\partial_{X_{1}}\,(\partial_{U_{3}}\!\!\cdot\partial_{X_{12}})^{2}\,,
	\label{332 exp}
\ea
but taking into account the symmetries under $1\leftrightarrow2$\,, one is left with four terms. One of such terms is \eqref{1 term}, and we have sketched how to get the corresponding (A)dS intrinsic expression \eqref{332 1 ds}. Applying the same techniques explained there, one can deal with the other three terms in the same manner.

We present the (A)dS intrinsic expression for each term. The third term in the expansion \eqref{332 exp} gives
\ba
	&& \partial_{U_{1}}\!\!\cdot\partial_{U_{3}}\,\partial_{U_{2}}\!\!\cdot\partial_{U_{3}}\,
	(\partial_{X_{2}}\!\!\cdot\partial_{U_{1}})^{2}\,
	(\partial_{X_{1}}\!\!\cdot\partial_{U_{2}})^{2} \nn
	 && \simeq\,
	 \partial_{u_{1}}\!\!\cdot\partial_{u_{3}}\,\partial_{u_{2}}\!\!\cdot\partial_{u_{3}}\,
	 (\partial_{u_{1}}\!\!\cdot D_{\sst 2})^{2} (\partial_{u_{2}}\!\!\cdot D_{\sst 1})^{2}
	 +\tfrac1{L^{2}}\,  \partial_{u_{1}}\!\!\cdot\partial_{u_{2}}\,
	 \partial_{u_{1}}\!\!\cdot\partial_{u_{3}}\,\partial_{u_{2}}\!\!\cdot\partial_{u_{3}}\,
	  \partial_{u_{1}}\!\!\cdot D_{\sst 2}\,\partial_{u_{2}}\!\!\cdot D_{\sst 1}
	 \nn
	 &&\quad\
	 -\,\tfrac1{L^{2}}\,\partial_{u_{1}}\!\!\cdot\partial_{u_{2}}\,
	 (\partial_{u_{1}}\!\!\cdot\partial_{u_{3}})^{2}\,(\partial_{u_{2}}\!\!\cdot D_{\sst 1})^{2}
	 -\tfrac1{L^{2}}\,\partial_{u_{1}}\!\!\cdot\partial_{u_{2}}\,
	 (\partial_{u_{2}}\!\!\cdot\partial_{u_{3}})^{2}\,(\partial_{u_{1}}\!\!\cdot D_{\sst 2})^{2}
	 \nn
	 &&\quad\
	+\,\tfrac{d+4}{L^{4}}\,(\partial_{u_{1}}\!\!\cdot\partial_{u_{2}})^{2}\,
	 \partial_{u_{1}}\!\!\cdot\partial_{u_{3}}\,\partial_{u_{2}}\!\!\cdot\partial_{u_{3}}\,,\qquad
\ea
where $\simeq$ means equivalence of two operators under the condition \eqref{hom cond}. The fourth term gives
\ba	
	&& \partial_{U_{1}}\!\!\cdot\partial_{U_{2}}\,\partial_{U_{2}}\!\!\cdot\partial_{U_{3}}\,
	(\partial_{U_{1}}\!\!\cdot\partial_{X_{2}})^{2}\,
	\partial_{U_{2}}\!\!\cdot\partial_{X_{1}}\,\partial_{U_{3}}\!\!\cdot\partial_{X_{12}} \nn
	&&\simeq\,
	\partial_{u_{1}}\!\!\cdot\partial_{u_{2}}\,\partial_{u_{2}}\!\!\cdot\partial_{u_{3}}\,
	(\partial_{u_{1}}\!\!\cdot D_{\sst 2})^{2}\,\partial_{u_{2}}\!\!\cdot D_{\sst 1}\,
	\partial_{u_{3}}\!\!\cdot D_{\sst 12}
	+\tfrac1{L^{2}}\,
	\partial_{u_{1}}\!\!\cdot\partial_{u_{2}}\,(\partial_{u_{2}}\!\!\cdot\partial_{u_{3}})^{2}
	(\partial_{u_{1}}\!\!\cdot D_{\sst 2})^{2} \nn
	&&\quad\
	-\tfrac1{L^{2}}\,(\partial_{u_{1}}\!\!\cdot\partial_{u_{2}})^{2}\,
	\partial_{u_{1}}\!\!\cdot\partial_{u_{3}}\,
	\partial_{u_{2}}\!\!\cdot D_{\sst 1}\,\partial_{u_{3}}\!\!\cdot D_{\sst 12}
	+\tfrac2{L^{2}}\,(\partial_{u_{1}}\!\!\cdot\partial_{u_{2}})^{2}\,
	\partial_{u_{2}}\!\!\cdot\partial_{u_{3}}\,
	\partial_{u_{1}}\!\!\cdot D_{\sst 2}\,\partial_{u_{3}}\!\!\cdot D_{\sst 12} \nn
	&&\quad\
	 +\tfrac3{L^{2}}\,\partial_{u_{1}}\!\!\cdot\partial_{u_{2}}\,
	 \partial_{u_{1}}\!\!\cdot\partial_{u_{3}}\,\partial_{u_{2}}\!\!\cdot\partial_{u_{3}}\,
	\,\partial_{u_{1}}\!\!\cdot D_{\sst 2}\,\partial_{u_{2}}\!\!\cdot D_{\sst 1}
	-\tfrac{d+1}{L^{4}}\,(\partial_{u_{1}}\!\!\cdot\partial_{u_{2}})^{2}\,
	 \partial_{u_{1}}\!\!\cdot\partial_{u_{3}}\,\partial_{u_{2}}\!\!\cdot\partial_{u_{3}}\,,\qquad
\ea
and the fifth term can be obtained interchanging 1 and 2 in the above. The last term gives
\ba
	&& (\partial_{U_{1}}\!\!\cdot\partial_{U_{2}})^{2}\,
	\partial_{U_{1}}\!\!\cdot\partial_{X_{2}}\,
	\partial_{U_{2}}\!\!\cdot\partial_{X_{1}}\,(\partial_{U_{3}}\!\!\cdot\partial_{X_{12}})^{2} \nn
	&&\simeq\,
	\partial_{u_{1}}\!\!\cdot D_{\sst 2}\,(\partial_{u_{2}}\!\!\cdot D_{\sst 12})^{2}
	+\tfrac2{L^{2}}\,(\partial_{u_{1}}\!\!\cdot\partial_{u_{2}})^{3}\,
	(\partial_{u_{3}}\!\!\cdot D_{\sst 12})^{2}\nn
	&&\quad\
	-\tfrac{5}{L^{2}}\,\partial_{u_{1}}\!\!\cdot\partial_{u_{3}}\,(\partial_{u_{1}}\!\!\cdot\partial_{u_{2}})^{2}\,
	\partial_{u_{2}}\!\!\cdot D_{\sst 1}\,\partial_{u_{3}}\!\!\cdot D_{\sst 2}
	-\tfrac{5}{L^{2}}\,\partial_{u_{2}}\!\!\cdot\partial_{u_{3}}\,(\partial_{u_{1}}\!\!\cdot\partial_{u_{2}})^{2}\,
	\partial_{u_{1}}\!\!\cdot D_{\sst 2}\,\partial_{u_{3}}\!\!\cdot D_{\sst 1} \nn
	&&\quad\
	-\tfrac8{L^{2}}\,\partial_{u_{1}}\!\!\cdot\partial_{u_{2}}\,
	 \partial_{u_{1}}\!\!\cdot\partial_{u_{3}}\,\partial_{u_{2}}\!\!\cdot\partial_{u_{3}}\,
	\partial_{u_{1}}\!\!\cdot D_{\sst 2}\,\partial_{u_{\sst 2}}\!\!\cdot D_{\sst 1}
	+2\,\tfrac{d-9}{L^{4}}\,(\partial_{u_{1}}\!\!\cdot\partial_{u_{2}})^{2}\,
	 \partial_{u_{1}}\!\!\cdot\partial_{u_{3}}\,\partial_{u_{2}}\!\!\cdot\partial_{u_{3}}\,,\qquad
\ea
and collecting all these terms finally gives \eqref{ex 332 dS}\,.

\bibliographystyle{JHEP}
\bibliography{ref}

\providecommand{\href}[2]{#2}\begingroup\raggedright\begin{thebibliography}{10}

\bibitem{SolvayHS}
R.~Argurio, G.~Barnich, G.~Bonelli, and M.~Grigoriev, eds., {\em ``Higher-Spin
  Gauge Theories'', Proceedings of the First Solvay Workshop, held in Brussels
  on May 12-14, 2004}, Int. Solvay Institutes, 2006.

\bibitem{Bianchi:2005yh}
M.~Bianchi and V.~Didenko, {\it {Massive higher spin multiplets and
  holography}},  \href{http://xxx.lanl.gov/abs/hep-th/0502220}{{\tt
  hep-th/0502220}}.

\bibitem{Francia:2005bv}
D.~Francia and C.~Hull, {\it {Higher-spin gauge fields and duality}},
  \href{http://xxx.lanl.gov/abs/hep-th/0501236}{{\tt hep-th/0501236}}.

\bibitem{Bouatta:2004kk}
N.~Bouatta, G.~Compere, and A.~Sagnotti, {\it {An Introduction to free
  higher-spin fields}},  \href{http://xxx.lanl.gov/abs/hep-th/0409068}{{\tt
  hep-th/0409068}}.

\bibitem{Bekaert:2005vh}
X.~Bekaert, S.~Cnockaert, C.~Iazeolla, and M.~Vasiliev, {\it {Nonlinear higher
  spin theories in various dimensions}},
  \href{http://xxx.lanl.gov/abs/hep-th/0503128}{{\tt hep-th/0503128}}.

\bibitem{Sagnotti:2005ns}
A.~Sagnotti, E.~Sezgin, and P.~Sundell, {\it {On higher spins with a strong
  Sp(2,R) condition}},  \href{http://xxx.lanl.gov/abs/hep-th/0501156}{{\tt
  hep-th/0501156}}.

\bibitem{Sorokin:2004ie}
D.~Sorokin, {\it {Introduction to the classical theory of higher spins}},  {\em
  AIP Conf.Proc.} {\bf 767} (2005) 172--202,
  [\href{http://xxx.lanl.gov/abs/hep-th/0405069}{{\tt hep-th/0405069}}].

\bibitem{Francia:2006hp}
D.~Francia and A.~Sagnotti, {\it {Higher-spin geometry and string theory}},
  {\em J.\ Phys.\ Conf.\ Ser.} {\bf 33} (2006) 57,
  [\href{http://xxx.lanl.gov/abs/hep-th/0601199}{{\tt hep-th/0601199}}].

\bibitem{Bekaert:2010hw}
X.~Bekaert, N.~Boulanger, and P.~Sundell, {\it {How higher-spin gravity
  surpasses the spin two barrier: no-go theorems versus yes-go examples}},
  \href{http://xxx.lanl.gov/abs/1007.0435}{{\tt arXiv:1007.0435}}.

\bibitem{Vasiliev:1988sa}
M.~A. Vasiliev, {\it {Consistent equations for interacting massless fields of
  all spins in the first order in curvatures}},  {\em Annals Phys.} {\bf 190}
  (1989) 59--106.

\bibitem{Vasiliev:2003ev}
M.~Vasiliev, {\it {Nonlinear equations for symmetric massless higher spin
  fields in (A)dS(d)}},  {\em Phys.Lett.} {\bf B567} (2003) 139--151,
  [\href{http://xxx.lanl.gov/abs/hep-th/0304049}{{\tt hep-th/0304049}}].

\bibitem{Vasiliev:2004qz}
M.~Vasiliev, {\it {Higher spin gauge theories in various dimensions}},  {\em
  Fortsch.\ Phys.} {\bf 52} (2004) 702--717,
  [\href{http://xxx.lanl.gov/abs/hep-th/0401177}{{\tt hep-th/0401177}}].

\bibitem{Boulanger:2011dd}
N.~Boulanger and P.~Sundell, {\it {An action principle for Vasiliev's
  four-dimensional higher-spin gravity}},
  \href{http://xxx.lanl.gov/abs/1102.2219}{{\tt arXiv:1102.2219}}.

\bibitem{Colombo:2010fu}
N.~Colombo and P.~Sundell, {\it {Twistor space observables and quasi-amplitudes
  in 4D higher spin gravity}},  \href{http://xxx.lanl.gov/abs/1012.0813}{{\tt
  arXiv:1012.0813}}.

\bibitem{Sezgin:2011hq}
E.~Sezgin and P.~Sundell, {\it {Geometry and Observables in Vasiliev's Higher
  Spin Gravity}},  \href{http://xxx.lanl.gov/abs/1103.2360}{{\tt
  arXiv:1103.2360}}.

\bibitem{Didenko:2008va}
V.~Didenko, A.~Matveev, and M.~Vasiliev, {\it {Unfolded Description of AdS(4)
  Kerr Black Hole}},  {\em Phys.Lett.} {\bf B665} (2008) 284--293,
  [\href{http://xxx.lanl.gov/abs/0801.2213}{{\tt arXiv:0801.2213}}].

\bibitem{Iazeolla:2008bp}
C.~Iazeolla, {\it {On the Algebraic Structure of Higher-Spin Field Equations
  and New Exact Solutions}},  \href{http://xxx.lanl.gov/abs/0807.0406}{{\tt
  arXiv:0807.0406}}.

\bibitem{Iazeolla:2011cb}
C.~Iazeolla and P.~Sundell, {\it {Families of exact solutions to Vasiliev's 4D
  equations with spherical, cylindrical and biaxial symmetry}},
  \href{http://xxx.lanl.gov/abs/1107.1217}{{\tt arXiv:1107.1217}}.

\bibitem{Fronsdal:1978rb}
C.~Fronsdal, {\it {Massless Fields with Integer Spin}},  {\em Phys.\ Rev.} {\bf
  D18} (1978) 3624.

\bibitem{Fang:1978wz}
J.~Fang and C.~Fronsdal, {\it {Massless Fields with Half Integral Spin}},  {\em
  Phys.\ Rev.} {\bf D18} (1978) 3630.

\bibitem{Francia:2002aa}
D.~Francia and A.~Sagnotti, {\it {Free geometric equations for higher spins}},
  {\em Phys.\ Lett.} {\bf B543} (2002) 303--310,
  [\href{http://xxx.lanl.gov/abs/hep-th/0207002}{{\tt hep-th/0207002}}].

\bibitem{Francia:2002pt}
D.~Francia and A.~Sagnotti, {\it {On the geometry of higher spin gauge
  fields}},  {\em Class.\ Quant.\ Grav.} {\bf 20} (2003) S473--S486,
  [\href{http://xxx.lanl.gov/abs/hep-th/0212185}{{\tt hep-th/0212185}}].

\bibitem{Francia:2005bu}
D.~Francia and A.~Sagnotti, {\it {Minimal local Lagrangians for higher-spin
  geometry}},  {\em Phys.\ Lett.} {\bf B624} (2005) 93--104,
  [\href{http://xxx.lanl.gov/abs/hep-th/0507144}{{\tt hep-th/0507144}}].

\bibitem{Campoleoni:2008jq}
A.~Campoleoni, D.~Francia, J.~Mourad, and A.~Sagnotti, {\it {Unconstrained
  Higher Spins of Mixed Symmetry. I. Bose Fields}},  {\em Nucl.\ Phys.} {\bf
  B815} (2009) 289--367, [\href{http://xxx.lanl.gov/abs/0810.4350}{{\tt
  arXiv:0810.4350}}].

\bibitem{Campoleoni:2009gs}
A.~Campoleoni, D.~Francia, J.~Mourad, and A.~Sagnotti, {\it {Unconstrained
  Higher Spins of Mixed Symmetry. II. Fermi Fields}},  {\em Nucl.\ Phys.} {\bf
  B828} (2010) 425, [\href{http://xxx.lanl.gov/abs/0904.4447}{{\tt
  arXiv:0904.4447}}].

\bibitem{Campoleoni:2009hj}
A.~Campoleoni, {\it {Lagrangian formulations for Bose and Fermi higher-spin
  fields of mixed symmetry}},  \href{http://xxx.lanl.gov/abs/0905.1472}{{\tt
  arXiv:0905.1472}}.

\bibitem{Campoleoni:2009je}
A.~Campoleoni, {\it {Metric-like Lagrangian Formulations for Higher-Spin Fields
  of Mixed Symmetry}},  {\em Riv.Nuovo Cim.} {\bf 033} (2010) 123--253,
  [\href{http://xxx.lanl.gov/abs/0910.3155}{{\tt arXiv:0910.3155}}].

\bibitem{Francia:2010ap}
D.~Francia, {\it {On the relation between local and geometric Lagrangians for
  higher spins}},  {\em J.Phys.Conf.Ser.} {\bf 222} (2010) 012002,
  [\href{http://xxx.lanl.gov/abs/1001.3854}{{\tt arXiv:1001.3854}}].

\bibitem{Francia:2010qp}
D.~Francia, {\it {String theory triplets and higher-spin curvatures}},  {\em
  Phys.Lett.} {\bf B690} (2010) 90--95,
  [\href{http://xxx.lanl.gov/abs/1001.5003}{{\tt arXiv:1001.5003}}].

\bibitem{Francia:2011qa}
D.~Francia, {\it {Low-spin models for higher-spin Lagrangians}},  {\em
  Prog.Theor.Phys.Suppl.} {\bf 188} (2011) 94--105,
  [\href{http://xxx.lanl.gov/abs/1103.0683}{{\tt arXiv:1103.0683}}].

\bibitem{Fradkin:1986qy}
E.~Fradkin and M.~A. Vasiliev, {\it {Cubic Interaction in Extended Theories of
  Massless Higher Spin Fields}},  {\em Nucl.\ Phys.} {\bf B291} (1987) 141.

\bibitem{Fradkin:1987ks}
E.~Fradkin and M.~A. Vasiliev, {\it {On the Gravitational Interaction of
  Massless Higher Spin Fields}},  {\em Phys.\ Lett.} {\bf B189} (1987) 89--95.

\bibitem{Vasiliev:2011xf}
M.~Vasiliev, {\it {Cubic Vertices for Symmetric Higher-Spin Gauge Fields in
  $(A)dS_d$}},  \href{http://xxx.lanl.gov/abs/1108.5921}{{\tt
  arXiv:1108.5921}}.

\bibitem{Alkalaev:2010af}
K.~Alkalaev, {\it {FV-type action for $AdS_5$ mixed-symmetry fields}},  {\em
  JHEP} {\bf 1103} (2011) 031, [\href{http://xxx.lanl.gov/abs/1011.6109}{{\tt
  arXiv:1011.6109}}].

\bibitem{Zinoviev:2010cr}
Y.~Zinoviev, {\it {Spin 3 cubic vertices in a frame-like formalism}},  {\em
  JHEP} {\bf 1008} (2010) 084, [\href{http://xxx.lanl.gov/abs/1007.0158}{{\tt
  arXiv:1007.0158}}].

\bibitem{Boulanger:2011qt}
N.~Boulanger, E.~Skvortsov, and Y.~Zinoviev, {\it {Gravitational cubic
  interactions for a simple mixed-symmetry gauge field in AdS and flat
  backgrounds}},  {\em J.Phys.A} {\bf A44} (2011) 415403,
  [\href{http://xxx.lanl.gov/abs/1107.1872}{{\tt arXiv:1107.1872}}].

\bibitem{Boulanger:2011se}
N.~Boulanger and E.~Skvortsov, {\it {Higher-spin algebras and cubic
  interactions for simple mixed-symmetry fields in AdS spacetime}},  {\em JHEP}
  {\bf 1109} (2011) 063, [\href{http://xxx.lanl.gov/abs/1107.5028}{{\tt
  arXiv:1107.5028}}].

\bibitem{Berends:1984rq}
F.~A. Berends, G.~Burgers, and H.~van Dam, {\it {On the theoretical problems in
  constructing interactions involving higher spin massless particles}},  {\em
  Nucl.\ Phys.} {\bf B260} (1985) 295.

\bibitem{Berends:1985xx}
F.~A. Berends, G.~Burgers, and H.~van Dam, {\it {Explicit construction of
  conserved currents for massless fields of arbitrary spin}},  {\em Nucl.\
  Phys.} {\bf B271} (1986) 429.

\bibitem{Fradkin:1995xy}
E.~Fradkin and R.~Metsaev, {\it {Cubic scattering amplitudes for all massless
  representations of the Poincare group in any space-time dimension}},  {\em
  Phys.Rev.} {\bf D52} (1995) 4660--4667.

\bibitem{Metsaev:2005ar}
R.~Metsaev, {\it {Cubic interaction vertices of massive and massless higher
  spin fields}},  {\em Nucl.Phys.} {\bf B759} (2006) 147--201,
  [\href{http://xxx.lanl.gov/abs/hep-th/0512342}{{\tt hep-th/0512342}}].

\bibitem{Metsaev:2007rn}
R.~Metsaev, {\it {Cubic interaction vertices for fermionic and bosonic
  arbitrary spin fields}},  \href{http://xxx.lanl.gov/abs/0712.3526}{{\tt
  arXiv:0712.3526}}.

\bibitem{Manvelyan:2010wp}
R.~Manvelyan, K.~Mkrtchyan, and W.~Ruehl, {\it {Direct construction of a cubic
  selfinteraction for higher spin gauge fields}},  {\em Nucl.Phys.} {\bf B844}
  (2011) 348--364, [\href{http://xxx.lanl.gov/abs/1002.1358}{{\tt
  arXiv:1002.1358}}].

\bibitem{Manvelyan:2010jr}
R.~Manvelyan, K.~Mkrtchyan, and W.~Ruehl, {\it {General trilinear interaction
  for arbitrary even higher spin gauge fields}},  {\em Nucl.\ Phys.} {\bf B836}
  (2010) 204--221, [\href{http://xxx.lanl.gov/abs/1003.2877}{{\tt
  arXiv:1003.2877}}].

\bibitem{Mkrtchyan:2010pp}
K.~Mkrtchyan, {\it {Higher Spin Interacting Quantum Field Theory and Higher
  Order Conformal Invariant Lagrangians}},
  \href{http://xxx.lanl.gov/abs/1011.0160}{{\tt arXiv:1011.0160}}. Ph.D.
  Thesis. Advisor Prof. Ruben Manvelyan.

\bibitem{Taronna:2010qq}
M.~Taronna, {\it {Higher Spins and String Interactions}},
  \href{http://xxx.lanl.gov/abs/1005.3061}{{\tt arXiv:1005.3061}}.

\bibitem{Sagnotti:2010at}
A.~Sagnotti and M.~Taronna, {\it {String Lessons for Higher-Spin
  Interactions}},  {\em Nucl. Phys.} {\bf B842} (2011) 299--361,
  [\href{http://xxx.lanl.gov/abs/1006.5242}{{\tt arXiv:1006.5242}}].

\bibitem{Taronna:2011kt}
M.~Taronna, {\it {Higher-Spin Interactions: four-point functions and beyond}},
  \href{http://xxx.lanl.gov/abs/1107.5843}{{\tt arXiv:1107.5843}}.

\bibitem{Fotopoulos:2010ay}
A.~Fotopoulos and M.~Tsulaia, {\it {On the Tensionless Limit of String theory,
  Off - Shell Higher Spin Interaction Vertices and BCFW Recursion Relations}},
  \href{http://xxx.lanl.gov/abs/1009.0727}{{\tt arXiv:1009.0727}}.

\bibitem{Zinoviev:2008ck}
Y.~M. Zinoviev, {\it {On spin 3 interacting with gravity}},  {\em Class. Quant.
  Grav.} {\bf 26} (2009) 035022, [\href{http://xxx.lanl.gov/abs/0805.2226}{{\tt
  arXiv:0805.2226}}].

\bibitem{Boulanger:2008tg}
N.~Boulanger, S.~Leclercq, and P.~Sundell, {\it {On The Uniqueness of Minimal
  Coupling in Higher-Spin Gauge Theory}},  {\em JHEP} {\bf 0808} (2008) 056,
  [\href{http://xxx.lanl.gov/abs/0805.2764}{{\tt arXiv:0805.2764}}].

\bibitem{Manvelyan:2009tf}
R.~Manvelyan and K.~Mkrtchyan, {\it {Conformal invariant interaction of a
  scalar field with the higher spin field in AdS(D)}},  {\em Mod.Phys.Lett.}
  {\bf A25} (2010) 1333--1348, [\href{http://xxx.lanl.gov/abs/0903.0058}{{\tt
  arXiv:0903.0058}}].

\bibitem{Fotopoulos:2007yq}
A.~Fotopoulos, N.~Irges, A.~C. Petkou, and M.~Tsulaia, {\it {Higher-Spin Gauge
  Fields Interacting with Scalars: The Lagrangian Cubic Vertex}},  {\em JHEP}
  {\bf 0710} (2007) 021, [\href{http://xxx.lanl.gov/abs/0708.1399}{{\tt
  arXiv:0708.1399}}].

\bibitem{Buchbinder:2006eq}
I.~Buchbinder, A.~Fotopoulos, A.~C. Petkou, and M.~Tsulaia, {\it {Constructing
  the cubic interaction vertex of higher spin gauge fields}},  {\em Phys.Rev.}
  {\bf D74} (2006) 105018, [\href{http://xxx.lanl.gov/abs/hep-th/0609082}{{\tt
  hep-th/0609082}}].

\bibitem{Fotopoulos:2008ka}
A.~Fotopoulos and M.~Tsulaia, {\it {Gauge Invariant Lagrangians for Free and
  Interacting Higher Spin Fields. A Review of the BRST formulation}},  {\em
  Int.J.Mod.Phys.} {\bf A24} (2009) 1--60,
  [\href{http://xxx.lanl.gov/abs/0805.1346}{{\tt arXiv:0805.1346}}]. Extended
  version of the contribution to the volume dedicated of Prof I.L. Buchbinder.

\bibitem{Fotopoulos:2010nj}
A.~Fotopoulos and M.~Tsulaia, {\it {Current Exchanges for Reducible Higher Spin
  Modes on AdS}},  \href{http://xxx.lanl.gov/abs/1007.0747}{{\tt
  arXiv:1007.0747}}.

\bibitem{Bekaert:2010hk}
X.~Bekaert and E.~Meunier, {\it {Higher spin interactions with scalar matter on
  constant curvature spacetimes: conserved current and cubic coupling
  generating functions}},  {\em JHEP} {\bf 11} (2010) 116,
  [\href{http://xxx.lanl.gov/abs/1007.4384}{{\tt arXiv:1007.4384}}].

\bibitem{Francia:2007qt}
D.~Francia, J.~Mourad, and A.~Sagnotti, {\it {Current Exchanges and
  Unconstrained Higher Spins}},  {\em Nucl.\ Phys.} {\bf B773} (2007) 203--237,
  [\href{http://xxx.lanl.gov/abs/hep-th/0701163}{{\tt hep-th/0701163}}].

\bibitem{Sagnotti:2010jt}
A.~Sagnotti, {\it {Higher Spins and Current Exchanges}},
  \href{http://xxx.lanl.gov/abs/1002.3388}{{\tt arXiv:1002.3388}}.

\bibitem{Bekaert:2009ud}
X.~Bekaert, E.~Joung, and J.~Mourad, {\it {On higher spin interactions with
  matter}},  {\em JHEP} {\bf 0905} (2009) 126,
  [\href{http://xxx.lanl.gov/abs/0903.3338}{{\tt arXiv:0903.3338}}].

\bibitem{Bekaert:2010ky}
X.~Bekaert, E.~Joung, and J.~Mourad, {\it {Effective action in a higher-spin
  background}},  {\em JHEP} {\bf 1102} (2011) 048,
  [\href{http://xxx.lanl.gov/abs/1012.2103}{{\tt arXiv:1012.2103}}].

\bibitem{Fronsdal:1978vb}
C.~Fronsdal, {\it {Singletons and Massless, Integral Spin Fields on de Sitter
  Space (Elementary Particles in a Curved Space. 7.)}},  {\em Phys.\ Rev.} {\bf
  D20} (1979) 848--856.

\bibitem{Manvelyan:2010je}
R.~Manvelyan, K.~Mkrtchyan, and W.~Ruehl, {\it {A generating function for the
  cubic interactions of higher spin fields}},  {\em Phys. Lett.} {\bf B696}
  (2011) 410--415, [\href{http://xxx.lanl.gov/abs/1009.1054}{{\tt
  arXiv:1009.1054}}].

\bibitem{Deser:1959zza}
S.~Deser and R.~Arnowitt, {\it {Quantum Theory of Gravitation I-Linearized
  Theory}},  {\em Phys.Rev.} {\bf 113} (1959) 745.

\bibitem{Metsaev:1995re}
R.~Metsaev, {\it {Massless mixed symmetry bosonic free fields in d-dimensional
  anti-de Sitter space-time}},  {\em Phys.Lett.} {\bf B354} (1995) 78--84.

\bibitem{Metsaev:1997nj}
R.~Metsaev, {\it {Arbitrary spin massless bosonic fields in d-dimensional
  anti-de Sitter space}},  \href{http://xxx.lanl.gov/abs/hep-th/9810231}{{\tt
  hep-th/9810231}}.

\bibitem{Biswas:2002nk}
T.~Biswas and W.~Siegel, {\it {Radial dimensional reduction: Anti-de Sitter
  theories from flat}},  {\em JHEP} {\bf 0207} (2002) 005,
  [\href{http://xxx.lanl.gov/abs/hep-th/0203115}{{\tt hep-th/0203115}}].

\bibitem{Bekaert:2003uc}
X.~Bekaert, I.~Buchbinder, A.~Pashnev, and M.~Tsulaia, {\it {On higher spin
  theory: Strings, BRST, dimensional reductions}},  {\em Class.\ Quant.\ Grav.}
  {\bf 21} (2004) S1457--1464,
  [\href{http://xxx.lanl.gov/abs/hep-th/0312252}{{\tt hep-th/0312252}}].

\bibitem{Hallowell:2005np}
K.~Hallowell and A.~Waldron, {\it {Constant curvature algebras and higher spin
  action generating functions}},  {\em Nucl.Phys.} {\bf B724} (2005) 453--486,
  [\href{http://xxx.lanl.gov/abs/hep-th/0505255}{{\tt hep-th/0505255}}].

\bibitem{Barnich:2006pc}
G.~Barnich and M.~Grigoriev, {\it {Parent form for higher spin fields on
  anti-de Sitter space}},  {\em JHEP} {\bf 0608} (2006) 013,
  [\href{http://xxx.lanl.gov/abs/hep-th/0602166}{{\tt hep-th/0602166}}].

\bibitem{Fotopoulos:2006ci}
A.~Fotopoulos, K.~L. Panigrahi, and M.~Tsulaia, {\it {Lagrangian formulation of
  higher spin theories on AdS space}},  {\em Phys.\ Rev.} {\bf D74} (2006)
  085029, [\href{http://xxx.lanl.gov/abs/hep-th/0607248}{{\tt
  hep-th/0607248}}].

\bibitem{Francia:2008hd}
D.~Francia, J.~Mourad, and A.~Sagnotti, {\it {(A)dS exchanges and
  partially-massless higher spins}},  {\em Nucl.\ Phys.} {\bf B804} (2008)
  383--420, [\href{http://xxx.lanl.gov/abs/0803.3832}{{\tt arXiv:0803.3832}}].

\bibitem{Boulanger:2008up}
N.~Boulanger, C.~Iazeolla, and P.~Sundell, {\it {Unfolding Mixed-Symmetry
  Fields in AdS and the BMV Conjecture: I. General Formalism}},  {\em JHEP}
  {\bf 0907} (2009) 013, [\href{http://xxx.lanl.gov/abs/0812.3615}{{\tt
  arXiv:0812.3615}}].

\bibitem{Boulanger:2008kw}
N.~Boulanger, C.~Iazeolla, and P.~Sundell, {\it {Unfolding Mixed-Symmetry
  Fields in AdS and the BMV Conjecture. II. Oscillator Realization}},  {\em
  JHEP} {\bf 0907} (2009) 014, [\href{http://xxx.lanl.gov/abs/0812.4438}{{\tt
  arXiv:0812.4438}}].

\bibitem{Alkalaev:2009vm}
K.~B. Alkalaev and M.~Grigoriev, {\it {Unified BRST description of AdS gauge
  fields}},  {\em Nucl.Phys.} {\bf B835} (2010) 197--220,
  [\href{http://xxx.lanl.gov/abs/0910.2690}{{\tt arXiv:0910.2690}}].

\bibitem{Boulanger:2005br}
N.~Boulanger, S.~Leclercq, and S.~Cnockaert, {\it {Parity violating vertices
  for spin-3 gauge fields}},  {\em Phys.\ Rev.} {\bf D73} (2006) 065019,
  [\href{http://xxx.lanl.gov/abs/hep-th/0509118}{{\tt hep-th/0509118}}].

\bibitem{Sagnotti:talk09}
A.~Sagnotti, {\it The higher-spin challenge},  in {\em Strings 2009}, 2009.
\newblock
  [\href{http://strings2009.roma2.infn.it/talks/Sagnotti_Strings09.pdf}{\tt
  http://strings2009.roma2.infn.it}].

\bibitem{Bekaert:2010hp}
X.~Bekaert, N.~Boulanger, and S.~Leclercq, {\it {Strong obstruction of the
  Berends-Burgers-van Dam spin-3 vertex}},  {\em J.\ Phys.\ A} {\bf A43} (2010)
  185401, [\href{http://xxx.lanl.gov/abs/1002.0289}{{\tt arXiv:1002.0289}}].

\bibitem{Manvelyan:2010jf}
R.~Manvelyan, K.~Mkrtchyan, W.~Ruhl, and M.~Tovmasyan, {\it {On Nonlinear
  Higher Spin Curvature}},  {\em Phys.Lett.} {\bf B699} (2011) 187--191,
  [\href{http://xxx.lanl.gov/abs/1102.0306}{{\tt arXiv:1102.0306}}].

\bibitem{Francia:2007ee}
D.~Francia, {\it {Geometric Lagrangians for massive higher-spin fields}},  {\em
  Nucl.\ Phys.} {\bf B796} (2008) 77--122,
  [\href{http://xxx.lanl.gov/abs/0710.5378}{{\tt arXiv:0710.5378}}].

\bibitem{Francia:2008ac}
D.~Francia, {\it {Geometric massive higher spins and current exchanges}},  {\em
  Fortsch.\ Phys.} {\bf 56} (2008) 800--808,
  [\href{http://xxx.lanl.gov/abs/0804.2857}{{\tt arXiv:0804.2857}}].

\bibitem{Zinoviev:2008jz}
Y.~Zinoviev, {\it {On spin 2 electromagnetic interactions}},  {\em
  Mod.Phys.Lett.} {\bf A24} (2009) 17--23,
  [\href{http://xxx.lanl.gov/abs/0806.4030}{{\tt arXiv:0806.4030}}].

\bibitem{Zinoviev:2009hu}
Y.~Zinoviev, {\it {On massive spin 2 electromagnetic interactions}},  {\em
  Nucl.Phys.} {\bf B821} (2009) 431--451,
  [\href{http://xxx.lanl.gov/abs/0901.3462}{{\tt arXiv:0901.3462}}].

\bibitem{Bengtsson:1986ys}
A.~K. Bengtsson, {\it {A Unified Action For Higher Spin Gauge Bosons From
  Covariant String Theory}},  {\em Phys.Lett.} {\bf B182} (1986) 321.

\bibitem{Heanault:book}
M.~Henneaux and C.~Teitelboim, {\em Quantum Mechanics of Fundamental System,
  2}.
\newblock Plenum Press, Newyork, 1987.

\bibitem{Sagnotti:2003qa}
A.~Sagnotti and M.~Tsulaia, {\it {On higher spins and the tensionless limit of
  string theory}},  {\em Nucl.Phys.} {\bf B682} (2004) 83--116,
  [\href{http://xxx.lanl.gov/abs/hep-th/0311257}{{\tt hep-th/0311257}}].

\bibitem{Porrati:2010hm}
M.~Porrati, R.~Rahman, and A.~Sagnotti, {\it {String Theory and The
  Velo-Zwanziger Problem}},  {\em Nucl.Phys.} {\bf B846} (2011) 250--282,
  [\href{http://xxx.lanl.gov/abs/1011.6411}{{\tt arXiv:1011.6411}}].

\bibitem{Polyakov:2009pk}
D.~Polyakov, {\it {Interactions of Massless Higher Spin Fields From String
  Theory}},  {\em Phys.\ Rev.} {\bf D82} (2010) 066005,
  [\href{http://xxx.lanl.gov/abs/0910.5338}{{\tt arXiv:0910.5338}}].

\bibitem{Polyakov:2010qs}
D.~Polyakov, {\it {Gravitational Couplings of Higher Spins from String
  Theory}},  {\em Int.\ J.\ Mod.\ Phys.} {\bf A25} (2010) 4623--4640,
  [\href{http://xxx.lanl.gov/abs/1005.5512}{{\tt arXiv:1005.5512}}].

\bibitem{Polyakov:2010sk}
D.~Polyakov, {\it {Higher Spins and Open Strings: Quartic Interactions}},  {\em
  Phys.Rev.} {\bf D83} (2011) 046005,
  [\href{http://xxx.lanl.gov/abs/1011.0353}{{\tt arXiv:1011.0353}}].

\bibitem{Polyakov:2011sm}
D.~Polyakov, {\it {A String Model for AdS Gravity and Higher Spins}},
  \href{http://xxx.lanl.gov/abs/1106.1558}{{\tt arXiv:1106.1558}}.

\bibitem{Sezgin:2002rt}
E.~Sezgin and P.~Sundell, {\it {Massless higher spins and holography}},  {\em
  Nucl.\ Phys.} {\bf B644} (2002) 303--370,
  [\href{http://xxx.lanl.gov/abs/hep-th/0205131}{{\tt hep-th/0205131}}].

\bibitem{Klebanov:2002ja}
I.~Klebanov and A.~Polyakov, {\it {AdS dual of the critical O(N) vector
  model}},  {\em Phys.\ Lett.} {\bf B550} (2002) 213--219,
  [\href{http://xxx.lanl.gov/abs/hep-th/0210114}{{\tt hep-th/0210114}}].

\bibitem{Campoleoni:2010zq}
A.~Campoleoni, S.~Fredenhagen, S.~Pfenninger, and S.~Theisen, {\it {Asymptotic
  symmetries of three-dimensional gravity coupled to higher-spin fields}},
  {\em JHEP} {\bf 1011} (2010) 007,
  [\href{http://xxx.lanl.gov/abs/1008.4744}{{\tt arXiv:1008.4744}}].

\bibitem{Henneaux:2010xg}
M.~Henneaux and S.-J. Rey, {\it {Nonlinear $W_{infinity}$ as Asymptotic
  Symmetry of Three-Dimensional Higher Spin Anti-de Sitter Gravity}},  {\em
  JHEP} {\bf 1012} (2010) 007, [\href{http://xxx.lanl.gov/abs/1008.4579}{{\tt
  arXiv:1008.4579}}].

\end{thebibliography}\endgroup

\end{document}